\documentclass[prl,reprint,superscriptaddress,noeprint,showkeys]{revtex4-2}

\usepackage[hidelinks]{hyperref}
\usepackage[dvipsnames]{xcolor}
\hypersetup{
    colorlinks,
    linkcolor={blue},
    citecolor={blue},
    urlcolor={blue},
}
\usepackage{graphicx}
\usepackage{amsmath,braket,graphicx,mathtools,amssymb}
\usepackage{amsfonts}
\usepackage{braket}
\usepackage{bm}

\newcommand{\RL}{$\rightarrow{}\!\!\!\leftarrow{}\;$\hspace{-0.5 em}}
\newcommand{\LR}{$\leftarrow{}\!\!\!\rightarrow{}\;$\hspace{-0.5 em}}
\newcommand{\RR}{$\rightarrow{}\!\!\!\rightarrow{}\;$\hspace{-0.5 em}}

\newcommand{\Si}{Si/SiO$_2$\space}

\begin{document}


\title{Space-Charge-Limited van der Waals Spin Transistor}


\author{Thomas K. M. Graham}

\author{Yu-Xuan Wang}
\affiliation{Department of Physics, Boston College, Chestnut Hill, MA 02467, USA}

\author{Niranjana Renjith Nair}
\author{Kseniia Mosina}
\affiliation{Department of Inorganic Chemistry, University of Chemistry and Technology Prague, Technick\'a 5, 166 28, Prague 6, Czechia}

\author{Kenji Watanabe}
\affiliation{Research Center for Electronic and Optical Materials, National Institute for Materials Science, 1-1 Namiki, Tsukuba 305-0044, Japan}

\author{Takashi Taniguchi}
\affiliation{Research Center for Materials Nanoarchitectonics, National Institute for Materials Science,  1-1 Namiki, Tsukuba 305-0044, Japan}

\author{Zden\v{e}k Sofer}
\affiliation{Department of Inorganic Chemistry, University of Chemistry and Technology Prague, Technick\'a 5, 166 28, Prague 6, Czechia}

\author{Brian B. Zhou}
\email{brian.zhou@bc.edu}
\affiliation{Department of Physics, Boston College, Chestnut Hill, MA 02467, USA}

\date{\today}
\begin{abstract}

Integrating semiconducting and magnetic materials could combine transistor-like operation with nonvolatility and enable architectures such as logic-in-memory. Here, we employ correlated electrical transport and scanning nitrogen-vacancy center magnetic imaging to elucidate a spin transistor concept that amalgamates vertical and lateral hopping transport inside a 2D antiferromagnetic semiconductor, mechanistically distinct from vertical tunneling devices. Our device, based on a monolayer-bilayer junction in CrSBr, displays giant, gate-tunable magnetoresistance driven by the dual action of electrostatic doping on space-charge-limited lateral conduction and interlayer exchange coupling. Moreover, we visualize a field-trainable, layer-sharing effect that selects between coherent or domain-wall reversal at the spin-flip transition, enabling multilevel, memristive conductance states. Our layer-dependent space charge mechanism for convergent electrical and magnetic control opens opportunities to address limitations in contemporary computing.
\end{abstract}



\maketitle

The von Neumann bottleneck, arising from data transfer between separate memory and logic units, is a central limitation to the efficiency of conventional computing. Spin transistors \cite{Datta1990,Sugahara2010,Tanaka2020} that colocalize nonvolatile memory and electrical switching could reduce system footprint and power dissipation during data transfer, while enabling reconfigurable logic gates \cite{Tanaka2007} and energy-efficient, ``normally-off" computing \cite{Ando2014}. Harnessing magnetic textures could unlock additional paradigms, including neuromorphic networks with synaptic weights encoded by the positions of domain walls in magnetic tunnel junctions \cite{Lequeux2016,Grollier2020a}.

In a spin transistor, a semiconducting channel connects magnetic source and drain contacts, with the device conductance modulated both electrically through an applied gate voltage and magnetically through the relative orientations of the source/drain magnetizations. Gate control can be achieved either by electric-field tuning of the Rashba spin-orbit interaction, which varies the spin precession within the channel (spin-FET \cite{Datta1990}), or by electrostatically modulating the channel's conductivity via carrier accumulation (spin-MOSFET \cite{Sugahara2010,Tanaka2007}). However, despite frontier materials advances \cite{Tanaka2020}, device performance remains limited by inefficient spin injection across interfaces \cite{Schmidt2000} and spin relaxation during transport \cite{Fabian1999}. 

Recently, layered antiferromagnetic (AFM) semiconductors with switchable interlayer spin configurations have figured prominently in spintronic device concepts \cite{Cardoso2018,Marian2023}. An archetype is the spin-filter magnetic tunnel junction (sf-MTJ), in which two graphene electrodes sandwich a few-layer AFM that provides a spin-dependent tunnel barrier \cite{Song2018a,Klein2018,Kim2018a,Wang2018a,Boix-Constant2024,Chen2024a,Liu2024a,Song2019,Jiang2019}. Notably, sf-MTJs made from CrI$_3$ have realized giant tunnel magnetoresistances (TMR) reaching 500\% for bilayers \cite{Song2018a} and a million percent in thicker multilayers \cite{Kim2018a} upon switching between antiparallel (AFM) and parallel (ferromagnetic, FM) alignment of consecutive layer magnetizations. Moreover, using the twist degree of freedom enables multilevel switching \cite{Boix-Constant2024} and bistability at zero field \cite{Chen2024a}.

\begin{figure*}[hbt]
\includegraphics[scale=1]{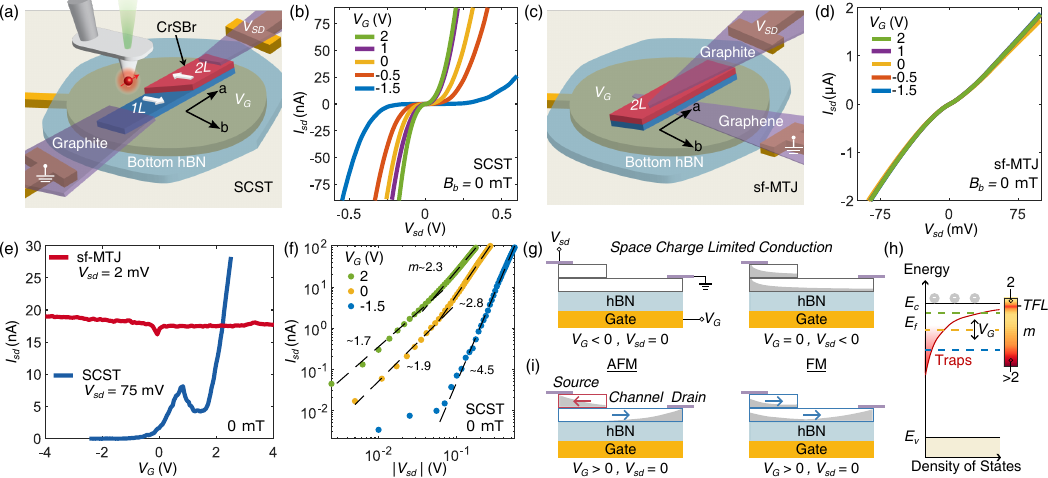}
\caption{\label{fig:1}Device overview. (a) Schematic of the SCST with contacts to two ends of a monolayer-bilayer step in CrSBr. $V_{sd}$ - bias, $V_{G}$ - gate. (b) Current $I_{sd}$ across the SCST versus bias $V_{sd}$ for several gates $V_G$ at zero field. (c) Schematic of an sf-MTJ. (d) $I_{sd}$ versus $V_{sd}$ for the sf-MTJ at zero field. (e) $I_{sd}$ versus $V_G$ at fixed bias for the SCST and sf-MTJ. (f) Log-log plot of $I_{sd}$ versus $V_{sd}$ for the SCST, showing a power law scaling $I_{sd} \propto V_{sd}^m$. Data for negative $V_{sd}$ is shown. (g) Schematic of SCL conduction. Left: Negative $V_G$ depletes the channel of free electrons. Right: Negative $V_{sd}$ injects electrons, forming a decaying distribution from the source contact. (h) Distribution of trap states, dominated by shallow defects, between the valence ($E_v$) and conduction ($E_c$) bands. The exponent $m$ initially decreases as the quasi-Fermi level $E_f$ rises. Upon reaching the trap-filled limit (TFL), $m$ rapidly increases and then decreases to two as $E_f$ crosses the conduction band. (i) Mechanism of MR in the SCST. Left: In the AFM state, electrons introduced through the contacts are primarily localized to the layer that each contact touches. Right: In the FM state, electrons distribute more uniformly between the two layers.}
\end{figure*}

Towards electrical control, experiments have exploited electrostatic doping to vary the strength of the AFM interlayer exchange coupling in CrI$_3$ and modulate the spin-filter conductance by switching between metamagnetic spin configurations \cite{Song2019,Jiang2019}. However, due to the limited range of exchange coupling modulation, reversible control requires operation in large magnetic fields near the metamagnetic transition. Alternatively, electrical control over the electrode's properties in the case of gapped bilayer graphene \cite{Jiang2019} and over interfacial effects \cite{Song2019} may also tune the tunnel conductance, but these mechanisms remain incompletely understood.

In this Letter, we demonstrate a conceptual shift where magnetoresistance (MR) in a layered AFM device derives not from changes in the tunnel barrier profile but from changes in the layer-dependent, space-charge density. Our device utilizes laterally separated contacts to different layers of a monolayer-bilayer step junction in CrSBr (Fig. \ref{fig:1}a), in contrast to sf-MTJs \cite{Song2018a,Klein2018,Kim2018a,Wang2018a,Boix-Constant2024,Chen2024a,Liu2024a,Song2019,Jiang2019} where tunneling between overlapping areas in the top and bottom electrodes determines the conductance (Fig. \ref{fig:1}c). This distinct approach, not possible for prior insulating CrI$_3$ \cite{Wang2018a}, amalgamates both vertical and lateral transport \emph{inside} the layered AFM. It allows both the interlayer magnetic ordering and an applied gate voltage to control the distribution of free carriers and thereby the conductance, realizing a spin transistor. By using a single material for spin polarization and transport, our design mitigates interfacial spin relaxation \cite{Datta1990,Sugahara2010,Tanaka2020,Schmidt2000,Fabian1999} and achieves a giant, gate-tunable TMR up to 3000\%.

Our understanding is informed by correlated electrical transport and magnetic imaging. We reveal that low-temperature transport in CrSBr \cite{Telford2022,Wu2022a} occurs in the trap-modulated, space-charge-limited (SCL) regime \cite{Lin2024,Mott1950,Rose1955,Mark1962,Lampert1964,Shi2015a}. Crucially, the resulting superlinear dependence on the bias voltage, with exponents tuned by electrostatic doping, enables higher MR than metallic spin valves in the ohmic regime \cite{Jedema2001}. Moreover, scanning nitrogen-vacancy (NV) center magnetometry \cite{Thiel2019, Song2021, Wang2025a, Pellet-Mary2025} visualizes how magnetic field training and gate-modulation of interlayer exchange coupling govern the formation and propagation of domain walls at the monolayer-bilayer interface, explaining the origin of partial steps and memory effects in the conductance.






\paragraph*{SCL conduction.---}Figure \ref{fig:1}a displays a schematic of our device, which we term space-charge-limited spin transistor (SCST), fabricated by van der Waals assembly (Appendix A). The source voltage $V_{sd}$ is applied to the top layer of the bilayer via a few-layer graphene contact (purple), while the current $I_{sd}$ flows into the grounded monolayer side predominantly along the $a$-axis of CrSBr. As the resistivity along the $a$-axis, reflecting hopping across incoherently coupled one-dimensional Cr-S chains, is two to five orders higher than along the $b$-axis \cite{Wu2022a}, our device can operate in a SCL regime, dictated by the build-up and self-repulsion of injected charges from the contact \cite{Mott1950,Rose1955,Mark1962,Lampert1964,Shi2015a,Lin2024}.


In Fig. \ref{fig:1}b, we present the $I_{sd}$ versus $V_{sd}$ curves for several gate voltages $V_G$ at zero external field and $T = 2$~K. Here, the bilayer possesses antiparallel layer magnetizations along the $b$-axis, the magnetic easy axis (e.g., Fig.~\ref{fig:1}a depicts the state \LR, denoting the magnetization of the top and bottom layer, respectively). For fixed $V_{sd}$, conduction is suppressed at negative $V_G$ (hole doping) and enhanced at positive $V_G$ (electron doping). This $n$-type conductance \cite{Weile2025} displays a pronounced nonlinearity versus $V_{sd}$, the signature of SCL current.

For comparison, we fabricate a conventional sf-MTJ using bilayer CrSBr (Fig.~\ref{fig:1}c). In contrast to the SCST, the $I_{sd}$-$V_{sd}$ curves for the sf-MTJ are nearly ohmic and display minimal dependence on $V_G$ (Fig. \ref{fig:1}d) \cite{Boix-Constant2022}. Fig. \ref{fig:1}e contrasts the gate dependences of the two architectures: while electron doping enhances the conductance of the SCST with an on/off ratio of $\sim$$10^4$, it barely affects the sf-MTJ. For a thin bilayer barrier, direct tunneling across the overlapping electrodes dominates the current, producing a linear dependence on $V_{sd}$ for small bias (see Supplemental Material (SM) \footnote{For the basic theory of SCL currents, details on the experimental setup and data analysis, and additional supporting data, see Supplemental Material at [url], which includes Refs. \cite{Gross2017,Dirnberger2023,Wang2017d}. Additional data include temperature and magnetic field angle dependence, alternative field training conditions, extended gate voltages, and a second SCST device.}). The sf-MTJ's conductance is largely insensitive to doping, since the barrier height for tunneling, determined by the alignment between the CrSBr conduction band and the Fermi level of the electrode, can be fixed by Fermi level pinning \cite{Liu2022b}.


In contrast, conduction in the SCST through the layered AFM is determined by its free carrier distribution $n = n_0 + n_g + n_{inj}$, given by the sum of the intrinsic $n_0$, gate-induced $n_g$ and bias-injected $n_{inj}$ free carrier densities. For metals, $n$ is dominated by $n_0$ (constant), and the bias determines the electric field $\mathbf{E}$, leading to ohmic conduction. However, for gapped materials with low $n_0$, $n$ can be instead dominated by $n_{inj}$, the uncompensated charge injected into the material due to an effect similar to capacitance. In this SCL regime, the distribution of $n_{inj}$ decays from the injecting contact (Fig. \ref{fig:1}g) and creates its own spatially varying $\mathbf{E}$, which self-consistently determines a continuous current \cite{Rose1955}. The bias thus regulates both $n$ and $\mathbf{E}$, causing a power-law $I_{sd}$-$V_{sd}$ scaling \cite{Note1}. For a thin film geometry absent charge traps (trap-free limit), the SCL sheet current density $K_{sd}$ is ideally quadratic: $K_{sd} = 2 \mu \varepsilon V_{sd}^2/(\pi L^2)$, where $\mu$ is the carrier mobility, $\varepsilon$ is the dielectric constant and $L$ is the channel length \cite{Geurst1966}.


In Fig. \ref{fig:1}f, we analyze the SCST's $I_{sd}$-$V_{sd}$ curves at zero field on a log-log plot. For large carrier injection at high bias, the curves display power-law scaling ($\propto V_{sd}^m$) with an exponent $m$ that diminishes from 4.5 to 2.3 with increasing electron doping. Exponents $m > 2$ indicate a trap-filling regime where a fraction of the injected charges populate nonconductive deep traps \cite{Rose1955,Mark1962,Lampert1964}. In this case, the quasi-Fermi level additionally rises with bias $V_{sd}$ (Fig.~\ref{fig:1}h), increasing the equilibrium ratio between conduction and trapped electrons. The current thus rises faster with bias ($m > 2$) as compared to the trap-free or trap-filled limits ($m = 2$), where all injected carriers are free. Specifically, larger $m$ corresponds to sweeping the quasi-Fermi level across intervals where the logarithmic derivative of the trap density of states with energy is smaller (i.e., flatter) \cite{Mark1962}. Hence, the reduced $m$ observed as the Fermi level rises with $V_G$ implies a trap distribution dominated by shallow donor states \cite{Weile2025}.

\begin{figure}[hbt]
\includegraphics[scale=1]{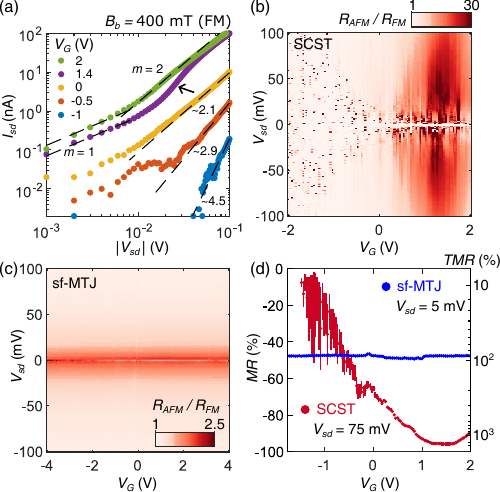}
\caption{\label{fig:2}Giant, gate-tunable MR in the SCST. (a) $I_{sd}$ versus $V_{sd}$ for the SCST in the FM state ($B_b = 0.4$~T) for select $V_G$. The $m=1$ and $m=2$ lines are guides. The arrow denotes the trap-filled transition. (b), (c) $R_{AFM}/R_{FM}$ versus $V_G$ and $V_{sd}$ for (b) the SCST and (c) the sf-MTJ at $T = 2$~K. (d) Linecuts of the magnetoresistance versus gate for the SCST and sf-MTJ at constant bias. The left and right $y$-axes denote different conventions MR and TMR, defined in the main text.}
\end{figure}


\paragraph*{Gate-tunable MR.---}
 The operation of our spin transistor relies on the sensitivity of the layer-resolved space charge distribution to both the magnetic state and gate voltage. At low temperature, $c$-axis conduction is mediated by variable range hopping between spin-polarized localized states aligned with their host layer magnetization \cite{Lin2024a}. In the AFM state, exchange splitting blocks electron hopping between adjacent layers, whereas no restriction occurs in the FM state. Thus, vertical hopping must overcome a spin-flip energy barrier in the AFM state. This activation energy flattens the $I_{sd}$-$V_{sd}$ curves at low bias and suppresses the linear transport regime to below our sensitivity (Fig. \ref{fig:1}f).

When electrons are accumulated into CrSBr by $V_G$ or $V_{sd}$, they should distribute differently in the AFM and FM states. Specifically, the added charges preferentially occupy the layer directly touching graphene in the AFM state (Fig. \ref{fig:1}i, left) \cite{Rizzo2025}, but diffuse between both layers in the FM state (Fig. \ref{fig:1}i, right). Since the SCST's conductance is limited by lateral conduction in the bottom monolayer, set by its free carrier distribution, MR at the AFM-to-FM transition should be strongly gate-tunable.

\begin{figure*}[hbt]
\includegraphics[scale=1]{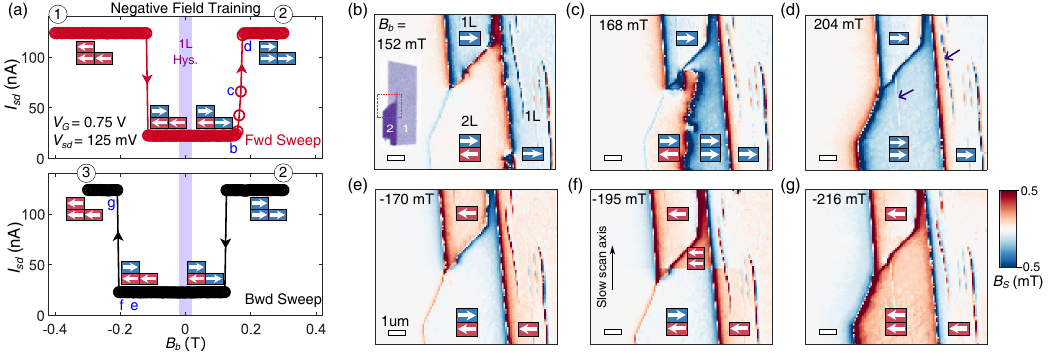}
\caption{\label{fig:3}Training effect on metamagnetic transitions. (a) Hysteresis sweep of the conductance under negative field training. The sweep starts at $B_b = -0.4$~T and pauses at an intermediate positive field (0.3 T) before ramping backward. Magnetic imaging determines the labeled bilayer/monolayer states. The monolayer's hysteresis loop is outlined by the purple box. (b,c,d) Magnetic imaging of the gradual AFM-to-FM transition on the forward sweep. The transition proceeds by reversing the bottom layer of the bilayer via domain-wall translation from the monolayer interface. Linear defects parallel to the $a$-axis (purple arrows) act as pinning sites. $B_s$ is the stray field along the NV axis. Optical image is shown as the inset in (b). (e,f,g) Images of the abrupt AFM-to-FM transition on the backward sweep, which instead occurs by a thermally-activated, coherent rotation of the top layer during scanning.}
\end{figure*}

 
Figure \ref{fig:2}a displays the $I_{sd}$-$V_{sd}$ curves of the SCST at an applied field $B_b = 400$~mT along the $b$-axis. Here, with the CrSBr bilayer in the FM state (\mbox{\RR)}, $I_{sd}$ is significantly enhanced compared to the AFM state (see Fig.~\ref{fig:5} for comparison on a linear scale). The higher conductance for the FM state results from more efficient charge injection into the bottom monolayer by the gate ($n_g$) and bias ($n_{inj}$), and by charge transfer from graphene to CrSBr ($n_0$) \cite{Rizzo2025}. This raises the quasi-Fermi level in the bottom layer, which manifests as a measurable ohmic current at low bias (dominated by $n_0 + n_g$) and smaller scaling exponents $m$ at high bias (dominated by $n_{inj}$) for the FM state. For $V_G > 1.6$~V, we observe a direct transition from ohmic to quadratic SCL conduction with increasing bias, indicating that the background carriers $n_g$ induced by $V_G$ essentially fill all trap states. Analyzing data at $V_G = 2$~V in the trap-filled regime, we estimate a mobility $\mu = 4000$~cm$^2$~V$^{-1}$~s$^{-1}$ for hBN-encapsulated CrSBr at 2~K \cite{Note1}, substantially higher than room temperature estimates (10~cm$^2$~V$^{-1}$~s$^{-1}$) \cite{Telford2022,Chou2025}. This increased mobility results primarily from the saturation of trap states, allowing transport to reflect the intrinsic nature of CrSBr's conduction bands \cite{Guo2018c}.

We summarize the ratio $R_{AFM}/R_{FM}$ between the resistances in the AFM and FM states versus gate and bias for the SCST and the sf-MTJ in Figs. \ref{fig:2}b and \ref{fig:2}c. To facilitate comparison, Fig. \ref{fig:2}d plots linecuts of the magnetoresistance at constant $V_{sd}$ against two $y$-axes: MR~$= (R_{FM} - R_{AFM})/R_{AFM}$ on the left and TMR = $(R_{AFM}-R_{FM})/R_{FM}$ on the right. The MR of the sf-MTJ is gate-independent \cite{Boix-Constant2022}, but decreases with bias from a low-bias value of $-50$\% (TMR = 100\%) \cite{Chen2024a}.

\begin{figure}[t]
\includegraphics[scale=1]{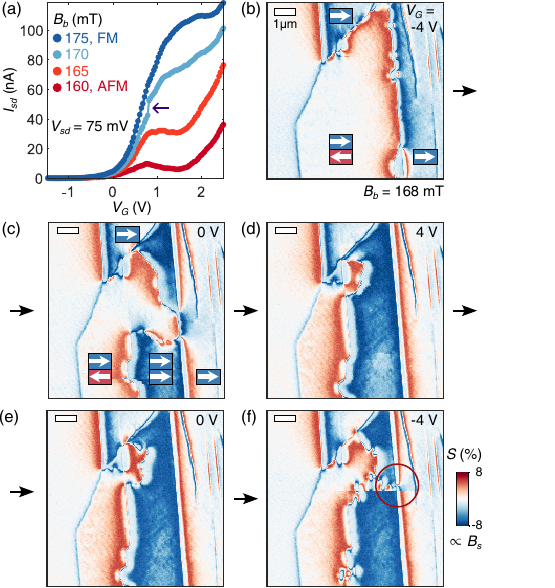}
\caption{\label{fig:4}Electrical modulation of interlayer exchange. (a) $I_{sd}$ versus $V_G$ at fields near the bilayer spin-flip transition. (b,c,d) Magnetic images for increasing gate voltage ($V_G = -4$, 0, and 4 V) at $B_b = 168$~mT. The FM regions of the bilayer expand with increasing electron doping. (e,f) Images as $V_G$ is decreased back to $-4$~V. The FM regions of the bilayer recede but are obstructed by pinning, introducing hysteresis.}
\end{figure}

In contrast, the MR of the SCST at 2 K is continuously tunable from $\sim$0 to $-97$\% (TMR = 3000\%) by increasing $V_G$ from $-2$ to 1.5 V. The vanishing MR for $V_G < 0$ results from electron depletion that shuts off conduction for both the AFM and FM states (see Fig. \ref{fig:4}a). The TMR is maximized around $|V_{sd}| \approx 30$~mV and $V_G \approx 1.4$~V when the conductivity of the FM state rapidly increases as carriers injected by the bias and gate voltages tune the channel towards the trap-filled limit (see arrow in Fig. \ref{fig:2}a). Here, the quasi-Fermi level rises quickly past diminishing trap states at the top of the distribution \cite{Lampert1964}.  

Moreover, the peak in the TMR at positive $V_G$ diminishes rapidly upon warming to 50~K (see Fig. S7 \cite{Note1}). Below 50 K, the conductances of the AFM and FM states follow temperature dependences consistent with variable-range hopping with different characteristic temperatures \cite{Shklovskii1984}; accordingly, the TMR displays a similar exponential decrease. At higher temperature, the SCST's TMR shows an additional rise for negative $V_G$ (\emph{reduced} carrier density), similar to that observed in lateral transport devices made from bulk CrSBr \cite{Chou2025}. The bulk MR is attributed to a magnetic state-dependent band edge shift and not directly sensitive to a single interlayer hopping as in our bilayer device. Hence, the SCST's crossover at low temperature to enhanced MR for higher carrier density highlights the pivotal role of nonlinear SCL transport and sharp, layer-dependent charge distributions due to suppressed lateral and vertical hopping.

\paragraph*{Correlative transport and magnetic imaging.---} We now leverage high-resolution magnetic imaging to elucidate the detailed origin of features observed in transport. We use $\langle100\rangle$-cut diamond tips with a nominal sensitivity of $1.9\;\mu \mathrm{T}/\sqrt{\mathrm{Hz}}$ and a standoff distance of $\approx$50 nm calibrated on our sample \cite{Note1}. For NV imaging, the magnetic field is oriented along the NV axis (35$^\circ$ out-of-plane). Due to strong magnetic anisotropy, the magnetization of CrSBr is determined mainly by $B_b$, the component along the $b$-axis, which we quote for comparison to transport measurements under $b$-axis only fields \cite{Note1}. Remarkably, the SCST's monolayer-bilayer interface provides cues to distinguish the antiphase states of the bilayer (\RL{} versus \LR{}), as shown in Appendix C. 



Figure \ref{fig:3}a displays the conductance hysteresis under a condition we term ``negative field training". Here, we begin at $B_b$ = $-0.4$~T, ramp to 0.3~T, before ramping back to $-0.3$~T. On the forward sweep, the FM-to-AFM transition for the bilayer at negative field is sharp, whereas multiple intermediate values of the conductance are observed for the AFM-to-FM transition on the positive side. Accordingly, NV imaging reveals that the FM-to-AFM transition occurs by a uniform reversal of the \emph{top} layer of the bilayer, avoiding the formation of a domain wall in the bottom layer and generating the state \RL{} (see labels in Fig. \ref{fig:3}a). Upon ramping the field positive, the monolayer first reverses from $\leftarrow{}$ to $\rightarrow$ at a small coercive field ($\sim$20~mT), creating a domain wall against the bilayer interface (Fig. \ref{fig:3}b). Thereafter, the AFM-to-FM transition occurs by translation of this domain wall across the bottom layer of the bilayer (Figs. \ref{fig:3}c), leading to fractional AFM and FM areas in the active region and intermediate values of the conductance.

In contrast, the backward sweep, starting from 0.3~T, displays two sharp conductance transitions. Here, NV imaging determines that the \emph{bottom} layer of the bilayer instead flips at the backward FM-to-AFM transition. Evidently, the bilayer retains a memory of its negative field training despite exhibiting saturated conductance for \RR. Notably, NV imaging cannot resolve any residual \RL{} domains within the field-of-view (Fig. \ref{fig:3}d). These observations suggest that nanoscale pinning sites, likely along the edge of CrSBr, remain antialigned even at 0.3 T. These remnants seed the abrupt reversal of the bottom layer as the field is reduced. Finally, after the monolayer flips to $\leftarrow$ at negative field (Fig. \ref{fig:3}e), only the top layer of the bilayer remains to be flipped to complete the AFM-to-FM transition, which occurs by a sudden rotation (Fig. \ref{fig:3}f). This understanding, where fields $\ge$0.4~T are needed to polarize the pinning sites and erase the training history, is corroborated by additional field sweeps \cite{Note1}.

\paragraph*{Gate-controlled interlayer exchange.---} Moreover, NV imaging extricates a second effect of gate voltage, hidden in transport. Figure \ref{fig:4}a presents $I_{sd}$ versus $V_G$ for several fields. The behaviors at $B_b$ = 160~mT and 175~mT are nearly identical to the full AFM and FM states, while intermediate conductances are observed at 165 mT and 170~mT, where an AFM-FM domain wall exists in the bilayer (Fig.~\ref{fig:3}). For both CrI$_3$ \cite{Huang2018,Jiang2018a} and CrSBr \cite{Tabataba-vakili2024}, electron doping is reported to suppress AFM interlayer exchange, lowering the spin-flip field to favor the high-conductance FM state. However, electron doping also increases the SCST's conductance by accumulating free carriers $n_g$ into the channel. Thus, the rising conductance with $V_G$ cannot isolate a change in interlayer magnetic order, although we occasionally detect discrete jumps that are incompatible with a continuous change in carrier concentration.

To determine whether doping modulates interlayer exchange, we image the SCST's active region at $B_b$ = 168~mT. Starting under strong electron depletion ($V_G = -4$~V, Fig. \ref{fig:4}b), the bilayer is nearly fully AFM (\RL), with its bottom layer forming a domain wall with the field-aligned monolayer ($\rightarrow$). As electron doping increases to $V_G$ = 4~V, the domain wall translates across the bilayer, progressively expanding the FM region (Fig. \ref{fig:4}c,d). Conversely, when $V_G$ is reduced back to $-4$~V, the FM domains in the bilayer retreat (Fig. \ref{fig:4}e,f). However, reversibility is hindered by strong pinning along vertical creases, which enables electrically-controlled memristance (Fig. \ref{fig:4}c and \ref{fig:4}e., both with gate off) \cite{Lequeux2016}. Abrupt Barkhausen jumps (circled in Fig. \ref{fig:4}f) are observed, consistent with the discontinuities seen in transport.

\paragraph*{Discussion.---} We demonstrated a distinct 2D spin transistor whose transport characteristics governed by layer-dependent space-charge accumulation is fundamentally distinct from traditional spin-filter tunneling. Our SCST accesses knobs unique to 2D magnetism, including gate control over the interlayer exchange and a layer-sharing effect that activates domain-wall control over the AFM-to-FM transition. Future efforts could optimize TMR by fabricating controlled multilayer geometries via laser cutting \cite{Sun2024} and improve the efficiency of electrical modulation by reducing pinning sites, likely introduced during device assembly here. These advances would position the SCST as a promising building block in nonvolatile or neuromorphic logic circuits.







\begin{thebibliography}{51}%
\makeatletter
\providecommand \@ifxundefined [1]{%
 \@ifx{#1\undefined}
}%
\providecommand \@ifnum [1]{%
 \ifnum #1\expandafter \@firstoftwo
 \else \expandafter \@secondoftwo
 \fi
}%
\providecommand \@ifx [1]{%
 \ifx #1\expandafter \@firstoftwo
 \else \expandafter \@secondoftwo
 \fi
}%
\providecommand \natexlab [1]{#1}%
\providecommand \enquote  [1]{``#1''}%
\providecommand \bibnamefont  [1]{#1}%
\providecommand \bibfnamefont [1]{#1}%
\providecommand \citenamefont [1]{#1}%
\providecommand \href@noop [0]{\@secondoftwo}%
\providecommand \href [0]{\begingroup \@sanitize@url \@href}%
\providecommand \@href[1]{\@@startlink{#1}\@@href}%
\providecommand \@@href[1]{\endgroup#1\@@endlink}%
\providecommand \@sanitize@url [0]{\catcode `\\12\catcode `\$12\catcode
  `\&12\catcode `\#12\catcode `\^12\catcode `\_12\catcode `\%12\relax}%
\providecommand \@@startlink[1]{}%
\providecommand \@@endlink[0]{}%
\providecommand \url  [0]{\begingroup\@sanitize@url \@url }%
\providecommand \@url [1]{\endgroup\@href {#1}{\urlprefix }}%
\providecommand \urlprefix  [0]{URL }%
\providecommand \Eprint [0]{\href }%
\providecommand \doibase [0]{https://doi.org/}%
\providecommand \selectlanguage [0]{\@gobble}%
\providecommand \bibinfo  [0]{\@secondoftwo}%
\providecommand \bibfield  [0]{\@secondoftwo}%
\providecommand \translation [1]{[#1]}%
\providecommand \BibitemOpen [0]{}%
\providecommand \bibitemStop [0]{}%
\providecommand \bibitemNoStop [0]{.\EOS\space}%
\providecommand \EOS [0]{\spacefactor3000\relax}%
\providecommand \BibitemShut  [1]{\csname bibitem#1\endcsname}%
\let\auto@bib@innerbib\@empty
\bibitem [{\citenamefont {Datta}\ and\ \citenamefont {Das}(1990)}]{Datta1990}%
  \BibitemOpen
  \bibfield  {author} {\bibinfo {author} {\bibfnamefont {S.}~\bibnamefont
  {Datta}}\ and\ \bibinfo {author} {\bibfnamefont {B.}~\bibnamefont {Das}},\
  }\bibfield  {title} {\bibinfo {title} {{Electronic analog of the
  electro-optic modulator}},\ }\href {https://doi.org/10.1063/1.102730}
  {\bibfield  {journal} {\bibinfo  {journal} {Applied Physics Letters}\
  }\textbf {\bibinfo {volume} {56}},\ \bibinfo {pages} {665} (\bibinfo {year}
  {1990})}\BibitemShut {NoStop}%
\bibitem [{\citenamefont {Sugahara}\ and\ \citenamefont
  {Nitta}(2010)}]{Sugahara2010}%
  \BibitemOpen
  \bibfield  {author} {\bibinfo {author} {\bibfnamefont {S.}~\bibnamefont
  {Sugahara}}\ and\ \bibinfo {author} {\bibfnamefont {J.}~\bibnamefont
  {Nitta}},\ }\bibfield  {title} {\bibinfo {title} {{Spin-transistor
  electronics: An overview and outlook}},\ }\href
  {https://doi.org/10.1109/JPROC.2010.2064272} {\bibfield  {journal} {\bibinfo
  {journal} {Proceedings of the IEEE}\ }\textbf {\bibinfo {volume} {98}},\
  \bibinfo {pages} {2124} (\bibinfo {year} {2010})}\BibitemShut {NoStop}%
\bibitem [{\citenamefont {Tanaka}(2021)}]{Tanaka2020}%
  \BibitemOpen
  \bibfield  {author} {\bibinfo {author} {\bibfnamefont {M.}~\bibnamefont
  {Tanaka}},\ }\bibfield  {title} {\bibinfo {title} {{Recent progress in
  ferromagnetic semiconductors and spintronics devices}},\ }\href
  {https://doi.org/10.35848/1347-4065/abcadc} {\bibfield  {journal} {\bibinfo
  {journal} {Japanese Journal of Applied Physics}\ }\textbf {\bibinfo {volume}
  {60}},\ \bibinfo {pages} {010101} (\bibinfo {year} {2021})}\BibitemShut
  {NoStop}%
\bibitem [{\citenamefont {Tanaka}\ and\ \citenamefont
  {Sugahara}(2007)}]{Tanaka2007}%
  \BibitemOpen
  \bibfield  {author} {\bibinfo {author} {\bibfnamefont {M.}~\bibnamefont
  {Tanaka}}\ and\ \bibinfo {author} {\bibfnamefont {S.}~\bibnamefont
  {Sugahara}},\ }\bibfield  {title} {\bibinfo {title} {{MOS-Based Spin Devices
  for Reconfigurable Logic}},\ }\href {https://doi.org/10.1109/TED.2007.894375}
  {\bibfield  {journal} {\bibinfo  {journal} {IEEE Transactions on Electron
  Devices}\ }\textbf {\bibinfo {volume} {54}},\ \bibinfo {pages} {961}
  (\bibinfo {year} {2007})}\BibitemShut {NoStop}%
\bibitem [{\citenamefont {Ando}\ \emph {et~al.}(2014)\citenamefont {Ando},
  \citenamefont {Fujita}, \citenamefont {Ito}, \citenamefont {Yuasa},
  \citenamefont {Suzuki}, \citenamefont {Nakatani}, \citenamefont {Miyazaki},\
  and\ \citenamefont {Yoda}}]{Ando2014}%
  \BibitemOpen
  \bibfield  {author} {\bibinfo {author} {\bibfnamefont {K.}~\bibnamefont
  {Ando}}, \bibinfo {author} {\bibfnamefont {S.}~\bibnamefont {Fujita}},
  \bibinfo {author} {\bibfnamefont {J.}~\bibnamefont {Ito}}, \bibinfo {author}
  {\bibfnamefont {S.}~\bibnamefont {Yuasa}}, \bibinfo {author} {\bibfnamefont
  {Y.}~\bibnamefont {Suzuki}}, \bibinfo {author} {\bibfnamefont
  {Y.}~\bibnamefont {Nakatani}}, \bibinfo {author} {\bibfnamefont
  {T.}~\bibnamefont {Miyazaki}},\ and\ \bibinfo {author} {\bibfnamefont
  {H.}~\bibnamefont {Yoda}},\ }\bibfield  {title} {\bibinfo {title}
  {Spin-transfer torque magnetoresistive random-access memory technologies for
  normally off computing (invited)},\ }\href
  {https://doi.org/10.1063/1.4869828} {\bibfield  {journal} {\bibinfo
  {journal} {Journal of Applied Physics}\ }\textbf {\bibinfo {volume} {115}},\
  \bibinfo {pages} {172607} (\bibinfo {year} {2014})}\BibitemShut {NoStop}%
\bibitem [{\citenamefont {Lequeux}\ \emph {et~al.}(2016)\citenamefont
  {Lequeux}, \citenamefont {Sampaio}, \citenamefont {Cros}, \citenamefont
  {Yakushiji}, \citenamefont {Fukushima}, \citenamefont {Matsumoto},
  \citenamefont {Kubota}, \citenamefont {Yuasa},\ and\ \citenamefont
  {Grollier}}]{Lequeux2016}%
  \BibitemOpen
  \bibfield  {author} {\bibinfo {author} {\bibfnamefont {S.}~\bibnamefont
  {Lequeux}}, \bibinfo {author} {\bibfnamefont {J.}~\bibnamefont {Sampaio}},
  \bibinfo {author} {\bibfnamefont {V.}~\bibnamefont {Cros}}, \bibinfo {author}
  {\bibfnamefont {K.}~\bibnamefont {Yakushiji}}, \bibinfo {author}
  {\bibfnamefont {A.}~\bibnamefont {Fukushima}}, \bibinfo {author}
  {\bibfnamefont {R.}~\bibnamefont {Matsumoto}}, \bibinfo {author}
  {\bibfnamefont {H.}~\bibnamefont {Kubota}}, \bibinfo {author} {\bibfnamefont
  {S.}~\bibnamefont {Yuasa}},\ and\ \bibinfo {author} {\bibfnamefont
  {J.}~\bibnamefont {Grollier}},\ }\bibfield  {title} {\bibinfo {title} {{A
  magnetic synapse: multilevel spin-torque memristor with perpendicular
  anisotropy}},\ }\href {https://doi.org/10.1038/srep31510} {\bibfield
  {journal} {\bibinfo  {journal} {Scientific Reports}\ }\textbf {\bibinfo
  {volume} {6}},\ \bibinfo {pages} {31510} (\bibinfo {year}
  {2016})}\BibitemShut {NoStop}%
\bibitem [{\citenamefont {Grollier}\ \emph {et~al.}(2020)\citenamefont
  {Grollier}, \citenamefont {Querlioz}, \citenamefont {Camsari}, \citenamefont
  {Everschor-Sitte}, \citenamefont {Fukami},\ and\ \citenamefont
  {Stiles}}]{Grollier2020a}%
  \BibitemOpen
  \bibfield  {author} {\bibinfo {author} {\bibfnamefont {J.}~\bibnamefont
  {Grollier}}, \bibinfo {author} {\bibfnamefont {D.}~\bibnamefont {Querlioz}},
  \bibinfo {author} {\bibfnamefont {K.~Y.}\ \bibnamefont {Camsari}}, \bibinfo
  {author} {\bibfnamefont {K.}~\bibnamefont {Everschor-Sitte}}, \bibinfo
  {author} {\bibfnamefont {S.}~\bibnamefont {Fukami}},\ and\ \bibinfo {author}
  {\bibfnamefont {M.~D.}\ \bibnamefont {Stiles}},\ }\bibfield  {title}
  {\bibinfo {title} {{Neuromorphic spintronics}},\ }\href
  {https://doi.org/10.1038/s41928-019-0360-9} {\bibfield  {journal} {\bibinfo
  {journal} {Nature Electronics}\ }\textbf {\bibinfo {volume} {3}},\ \bibinfo
  {pages} {360} (\bibinfo {year} {2020})}\BibitemShut {NoStop}%
\bibitem [{\citenamefont {Schmidt}\ \emph {et~al.}(2000)\citenamefont
  {Schmidt}, \citenamefont {Ferrand}, \citenamefont {Molenkamp}, \citenamefont
  {Filip},\ and\ \citenamefont {van Wees}}]{Schmidt2000}%
  \BibitemOpen
  \bibfield  {author} {\bibinfo {author} {\bibfnamefont {G.}~\bibnamefont
  {Schmidt}}, \bibinfo {author} {\bibfnamefont {D.}~\bibnamefont {Ferrand}},
  \bibinfo {author} {\bibfnamefont {L.~W.}\ \bibnamefont {Molenkamp}}, \bibinfo
  {author} {\bibfnamefont {A.~T.}\ \bibnamefont {Filip}},\ and\ \bibinfo
  {author} {\bibfnamefont {B.~J.}\ \bibnamefont {van Wees}},\ }\bibfield
  {title} {\bibinfo {title} {{Fundamental obstacle for electrical spin
  injection from a ferromagnetic metal into a diffusive semiconductor}},\
  }\href {https://doi.org/10.1103/PhysRevB.62.R4790} {\bibfield  {journal}
  {\bibinfo  {journal} {Physical Review B}\ }\textbf {\bibinfo {volume} {62}},\
  \bibinfo {pages} {R4790} (\bibinfo {year} {2000})}\BibitemShut {NoStop}%
\bibitem [{\citenamefont {Fabian}\ and\ \citenamefont
  {Sarma}(1999)}]{Fabian1999}%
  \BibitemOpen
  \bibfield  {author} {\bibinfo {author} {\bibfnamefont {J.}~\bibnamefont
  {Fabian}}\ and\ \bibinfo {author} {\bibfnamefont {S.~D.}\ \bibnamefont
  {Sarma}},\ }\bibfield  {title} {\bibinfo {title} {{Spin relaxation of
  conduction electrons}},\ }\href {https://doi.org/10.1116/1.590813} {\bibfield
   {journal} {\bibinfo  {journal} {J. Vac. Sci. Technol. B}\ }\textbf {\bibinfo
  {volume} {17}},\ \bibinfo {pages} {1708} (\bibinfo {year}
  {1999})}\BibitemShut {NoStop}%
\bibitem [{\citenamefont {Cardoso}\ \emph {et~al.}(2018)\citenamefont
  {Cardoso}, \citenamefont {Soriano}, \citenamefont
  {Garc{\'{i}}a-Mart{\'{i}}nez},\ and\ \citenamefont
  {Fern{\'{a}}ndez-Rossier}}]{Cardoso2018}%
  \BibitemOpen
  \bibfield  {author} {\bibinfo {author} {\bibfnamefont {C.}~\bibnamefont
  {Cardoso}}, \bibinfo {author} {\bibfnamefont {D.}~\bibnamefont {Soriano}},
  \bibinfo {author} {\bibfnamefont {N.~A.}\ \bibnamefont
  {Garc{\'{i}}a-Mart{\'{i}}nez}},\ and\ \bibinfo {author} {\bibfnamefont
  {J.}~\bibnamefont {Fern{\'{a}}ndez-Rossier}},\ }\bibfield  {title} {\bibinfo
  {title} {{Van der Waals Spin Valves}},\ }\href
  {https://doi.org/10.1103/PhysRevLett.121.067701} {\bibfield  {journal}
  {\bibinfo  {journal} {Physical Review Letters}\ }\textbf {\bibinfo {volume}
  {121}},\ \bibinfo {pages} {067701} (\bibinfo {year} {2018})}\BibitemShut
  {NoStop}%
\bibitem [{\citenamefont {Marian}\ \emph {et~al.}(2023)\citenamefont {Marian},
  \citenamefont {Soriano}, \citenamefont {Cannav{\'{o}}}, \citenamefont
  {Marin},\ and\ \citenamefont {Fiori}}]{Marian2023}%
  \BibitemOpen
  \bibfield  {author} {\bibinfo {author} {\bibfnamefont {D.}~\bibnamefont
  {Marian}}, \bibinfo {author} {\bibfnamefont {D.}~\bibnamefont {Soriano}},
  \bibinfo {author} {\bibfnamefont {E.}~\bibnamefont {Cannav{\'{o}}}}, \bibinfo
  {author} {\bibfnamefont {E.~G.}\ \bibnamefont {Marin}},\ and\ \bibinfo
  {author} {\bibfnamefont {G.}~\bibnamefont {Fiori}},\ }\bibfield  {title}
  {\bibinfo {title} {{Electrically tunable lateral spin-valve transistor based
  on bilayer CrI3}},\ }\href {https://doi.org/10.1038/s41699-023-00400-5}
  {\bibfield  {journal} {\bibinfo  {journal} {npj 2D Materials and
  Applications}\ }\textbf {\bibinfo {volume} {7}},\ \bibinfo {pages} {42}
  (\bibinfo {year} {2023})}\BibitemShut {NoStop}%
\bibitem [{\citenamefont {Song}\ \emph {et~al.}(2018)\citenamefont {Song},
  \citenamefont {Cai}, \citenamefont {Tu}, \citenamefont {Zhang}, \citenamefont
  {Huang}, \citenamefont {Wilson}, \citenamefont {Seyler}, \citenamefont {Zhu},
  \citenamefont {Taniguchi}, \citenamefont {Watanabe}, \citenamefont {McGuire},
  \citenamefont {Cobden}, \citenamefont {Xiao}, \citenamefont {Yao},\ and\
  \citenamefont {Xu}}]{Song2018a}%
  \BibitemOpen
  \bibfield  {author} {\bibinfo {author} {\bibfnamefont {T.}~\bibnamefont
  {Song}}, \bibinfo {author} {\bibfnamefont {X.}~\bibnamefont {Cai}}, \bibinfo
  {author} {\bibfnamefont {M.~W.-Y.}\ \bibnamefont {Tu}}, \bibinfo {author}
  {\bibfnamefont {X.}~\bibnamefont {Zhang}}, \bibinfo {author} {\bibfnamefont
  {B.}~\bibnamefont {Huang}}, \bibinfo {author} {\bibfnamefont {N.~P.}\
  \bibnamefont {Wilson}}, \bibinfo {author} {\bibfnamefont {K.~L.}\
  \bibnamefont {Seyler}}, \bibinfo {author} {\bibfnamefont {L.}~\bibnamefont
  {Zhu}}, \bibinfo {author} {\bibfnamefont {T.}~\bibnamefont {Taniguchi}},
  \bibinfo {author} {\bibfnamefont {K.}~\bibnamefont {Watanabe}}, \bibinfo
  {author} {\bibfnamefont {M.~A.}\ \bibnamefont {McGuire}}, \bibinfo {author}
  {\bibfnamefont {D.~H.}\ \bibnamefont {Cobden}}, \bibinfo {author}
  {\bibfnamefont {D.}~\bibnamefont {Xiao}}, \bibinfo {author} {\bibfnamefont
  {W.}~\bibnamefont {Yao}},\ and\ \bibinfo {author} {\bibfnamefont
  {X.}~\bibnamefont {Xu}},\ }\bibfield  {title} {\bibinfo {title} {{Giant
  tunneling magnetoresistance in spin-filter van der Waals heterostructures}},\
  }\href {https://doi.org/10.1126/science.aar4851} {\bibfield  {journal}
  {\bibinfo  {journal} {Science}\ }\textbf {\bibinfo {volume} {360}},\ \bibinfo
  {pages} {1214} (\bibinfo {year} {2018})}\BibitemShut {NoStop}%
\bibitem [{\citenamefont {Klein}\ \emph {et~al.}(2018)\citenamefont {Klein},
  \citenamefont {MacNeill}, \citenamefont {Lado}, \citenamefont {Soriano},
  \citenamefont {Navarro-Moratalla}, \citenamefont {Watanabe}, \citenamefont
  {Taniguchi}, \citenamefont {Manni}, \citenamefont {Canfield}, \citenamefont
  {Fern{\'{a}}ndez-Rossier},\ and\ \citenamefont
  {Jarillo-Herrero}}]{Klein2018}%
  \BibitemOpen
  \bibfield  {author} {\bibinfo {author} {\bibfnamefont {D.~R.}\ \bibnamefont
  {Klein}}, \bibinfo {author} {\bibfnamefont {D.}~\bibnamefont {MacNeill}},
  \bibinfo {author} {\bibfnamefont {J.~L.}\ \bibnamefont {Lado}}, \bibinfo
  {author} {\bibfnamefont {D.}~\bibnamefont {Soriano}}, \bibinfo {author}
  {\bibfnamefont {E.}~\bibnamefont {Navarro-Moratalla}}, \bibinfo {author}
  {\bibfnamefont {K.}~\bibnamefont {Watanabe}}, \bibinfo {author}
  {\bibfnamefont {T.}~\bibnamefont {Taniguchi}}, \bibinfo {author}
  {\bibfnamefont {S.}~\bibnamefont {Manni}}, \bibinfo {author} {\bibfnamefont
  {P.}~\bibnamefont {Canfield}}, \bibinfo {author} {\bibfnamefont
  {J.}~\bibnamefont {Fern{\'{a}}ndez-Rossier}},\ and\ \bibinfo {author}
  {\bibfnamefont {P.}~\bibnamefont {Jarillo-Herrero}},\ }\bibfield  {title}
  {\bibinfo {title} {{Probing magnetism in 2D van der Waals crystalline
  insulators via electron tunneling}},\ }\href
  {https://doi.org/10.1126/science.aar3617} {\bibfield  {journal} {\bibinfo
  {journal} {Science}\ }\textbf {\bibinfo {volume} {360}},\ \bibinfo {pages}
  {1218} (\bibinfo {year} {2018})}\BibitemShut {NoStop}%
\bibitem [{\citenamefont {Kim}\ \emph {et~al.}(2018)\citenamefont {Kim},
  \citenamefont {Yang}, \citenamefont {Patel}, \citenamefont {Sfigakis},
  \citenamefont {Li}, \citenamefont {Tian}, \citenamefont {Lei},\ and\
  \citenamefont {Tsen}}]{Kim2018a}%
  \BibitemOpen
  \bibfield  {author} {\bibinfo {author} {\bibfnamefont {H.~H.}\ \bibnamefont
  {Kim}}, \bibinfo {author} {\bibfnamefont {B.}~\bibnamefont {Yang}}, \bibinfo
  {author} {\bibfnamefont {T.}~\bibnamefont {Patel}}, \bibinfo {author}
  {\bibfnamefont {F.}~\bibnamefont {Sfigakis}}, \bibinfo {author}
  {\bibfnamefont {C.}~\bibnamefont {Li}}, \bibinfo {author} {\bibfnamefont
  {S.}~\bibnamefont {Tian}}, \bibinfo {author} {\bibfnamefont {H.}~\bibnamefont
  {Lei}},\ and\ \bibinfo {author} {\bibfnamefont {A.~W.}\ \bibnamefont
  {Tsen}},\ }\bibfield  {title} {\bibinfo {title} {{One Million Percent Tunnel
  Magnetoresistance in a Magnetic van der Waals Heterostructure}},\ }\href
  {https://doi.org/10.1021/acs.nanolett.8b01552} {\bibfield  {journal}
  {\bibinfo  {journal} {Nano Letters}\ }\textbf {\bibinfo {volume} {18}},\
  \bibinfo {pages} {4885} (\bibinfo {year} {2018})}\BibitemShut {NoStop}%
\bibitem [{\citenamefont {Wang}\ \emph {et~al.}(2018)\citenamefont {Wang},
  \citenamefont {Guti{\'{e}}rrez-Lezama}, \citenamefont {Ubrig}, \citenamefont
  {Kroner}, \citenamefont {Gibertini}, \citenamefont {Taniguchi}, \citenamefont
  {Watanabe}, \citenamefont {Imamoğlu}, \citenamefont {Giannini},\ and\
  \citenamefont {Morpurgo}}]{Wang2018a}%
  \BibitemOpen
  \bibfield  {author} {\bibinfo {author} {\bibfnamefont {Z.}~\bibnamefont
  {Wang}}, \bibinfo {author} {\bibfnamefont {I.}~\bibnamefont
  {Guti{\'{e}}rrez-Lezama}}, \bibinfo {author} {\bibfnamefont {N.}~\bibnamefont
  {Ubrig}}, \bibinfo {author} {\bibfnamefont {M.}~\bibnamefont {Kroner}},
  \bibinfo {author} {\bibfnamefont {M.}~\bibnamefont {Gibertini}}, \bibinfo
  {author} {\bibfnamefont {T.}~\bibnamefont {Taniguchi}}, \bibinfo {author}
  {\bibfnamefont {K.}~\bibnamefont {Watanabe}}, \bibinfo {author}
  {\bibfnamefont {A.}~\bibnamefont {Imamoğlu}}, \bibinfo {author}
  {\bibfnamefont {E.}~\bibnamefont {Giannini}},\ and\ \bibinfo {author}
  {\bibfnamefont {A.~F.}\ \bibnamefont {Morpurgo}},\ }\bibfield  {title}
  {\bibinfo {title} {{Very large tunneling magnetoresistance in layered
  magnetic semiconductor CrI3}},\ }\href
  {https://doi.org/10.1038/s41467-018-04953-8} {\bibfield  {journal} {\bibinfo
  {journal} {Nature Communications}\ }\textbf {\bibinfo {volume} {9}},\
  \bibinfo {pages} {2516} (\bibinfo {year} {2018})}\BibitemShut {NoStop}%
\bibitem [{\citenamefont {Boix-Constant}\ \emph {et~al.}(2024)\citenamefont
  {Boix-Constant}, \citenamefont {Jenkins}, \citenamefont {Rama-Eiroa},
  \citenamefont {Santos}, \citenamefont {Ma{\~{n}}as-Valero},\ and\
  \citenamefont {Coronado}}]{Boix-Constant2024}%
  \BibitemOpen
  \bibfield  {author} {\bibinfo {author} {\bibfnamefont {C.}~\bibnamefont
  {Boix-Constant}}, \bibinfo {author} {\bibfnamefont {S.}~\bibnamefont
  {Jenkins}}, \bibinfo {author} {\bibfnamefont {R.}~\bibnamefont {Rama-Eiroa}},
  \bibinfo {author} {\bibfnamefont {E.~J.~G.}\ \bibnamefont {Santos}}, \bibinfo
  {author} {\bibfnamefont {S.}~\bibnamefont {Ma{\~{n}}as-Valero}},\ and\
  \bibinfo {author} {\bibfnamefont {E.}~\bibnamefont {Coronado}},\ }\bibfield
  {title} {\bibinfo {title} {{Multistep magnetization switching in orthogonally
  twisted ferromagnetic monolayers}},\ }\href
  {https://doi.org/10.1038/s41563-023-01735-6} {\bibfield  {journal} {\bibinfo
  {journal} {Nature Materials}\ }\textbf {\bibinfo {volume} {23}},\ \bibinfo
  {pages} {212} (\bibinfo {year} {2024})}\BibitemShut {NoStop}%
\bibitem [{\citenamefont {Chen}\ \emph {et~al.}(2024)\citenamefont {Chen},
  \citenamefont {Samanta}, \citenamefont {Shahed}, \citenamefont {Zhang},
  \citenamefont {Fang}, \citenamefont {Ernst}, \citenamefont {Tsymbal},\ and\
  \citenamefont {Parkin}}]{Chen2024a}%
  \BibitemOpen
  \bibfield  {author} {\bibinfo {author} {\bibfnamefont {Y.}~\bibnamefont
  {Chen}}, \bibinfo {author} {\bibfnamefont {K.}~\bibnamefont {Samanta}},
  \bibinfo {author} {\bibfnamefont {N.~A.}\ \bibnamefont {Shahed}}, \bibinfo
  {author} {\bibfnamefont {H.}~\bibnamefont {Zhang}}, \bibinfo {author}
  {\bibfnamefont {C.}~\bibnamefont {Fang}}, \bibinfo {author} {\bibfnamefont
  {A.}~\bibnamefont {Ernst}}, \bibinfo {author} {\bibfnamefont {E.~Y.}\
  \bibnamefont {Tsymbal}},\ and\ \bibinfo {author} {\bibfnamefont {S.~S.~P.}\
  \bibnamefont {Parkin}},\ }\bibfield  {title} {\bibinfo {title}
  {{Twist-assisted all-antiferromagnetic tunnel junction in the atomic
  limit}},\ }\href {https://doi.org/10.1038/s41586-024-07818-x} {\bibfield
  {journal} {\bibinfo  {journal} {Nature}\ }\textbf {\bibinfo {volume} {632}},\
  \bibinfo {pages} {1045} (\bibinfo {year} {2024})}\BibitemShut {NoStop}%
\bibitem [{\citenamefont {Liu}\ \emph {et~al.}(2025)\citenamefont {Liu},
  \citenamefont {Sun}, \citenamefont {Zhu}, \citenamefont {Hong}, \citenamefont
  {Gao}, \citenamefont {Sun}, \citenamefont {Watanabe}, \citenamefont
  {Taniguchi}, \citenamefont {Wu}, \citenamefont {Chen}, \citenamefont {Gu},\
  and\ \citenamefont {Ye}}]{Liu2024a}%
  \BibitemOpen
  \bibfield  {author} {\bibinfo {author} {\bibfnamefont {Z.}~\bibnamefont
  {Liu}}, \bibinfo {author} {\bibfnamefont {Y.}~\bibnamefont {Sun}}, \bibinfo
  {author} {\bibfnamefont {C.}~\bibnamefont {Zhu}}, \bibinfo {author}
  {\bibfnamefont {C.}~\bibnamefont {Hong}}, \bibinfo {author} {\bibfnamefont
  {Y.}~\bibnamefont {Gao}}, \bibinfo {author} {\bibfnamefont {Z.}~\bibnamefont
  {Sun}}, \bibinfo {author} {\bibfnamefont {K.}~\bibnamefont {Watanabe}},
  \bibinfo {author} {\bibfnamefont {T.}~\bibnamefont {Taniguchi}}, \bibinfo
  {author} {\bibfnamefont {S.}~\bibnamefont {Wu}}, \bibinfo {author}
  {\bibfnamefont {Z.}~\bibnamefont {Chen}}, \bibinfo {author} {\bibfnamefont
  {P.}~\bibnamefont {Gu}},\ and\ \bibinfo {author} {\bibfnamefont
  {Y.}~\bibnamefont {Ye}},\ }\bibfield  {title} {\bibinfo {title} {{Spin
  texture and tunneling magnetoresistance in atomically thin CrSBr}},\ }\href
  {https://doi.org/10.1103/PhysRevB.111.L140417} {\bibfield  {journal}
  {\bibinfo  {journal} {Physical Review B}\ }\textbf {\bibinfo {volume}
  {111}},\ \bibinfo {pages} {L140417} (\bibinfo {year} {2025})}\BibitemShut
  {NoStop}%
\bibitem [{\citenamefont {Song}\ \emph {et~al.}(2019)\citenamefont {Song},
  \citenamefont {Tu}, \citenamefont {Carnahan}, \citenamefont {Cai},
  \citenamefont {Taniguchi}, \citenamefont {Watanabe}, \citenamefont {McGuire},
  \citenamefont {Cobden}, \citenamefont {Xiao}, \citenamefont {Yao},\ and\
  \citenamefont {Xu}}]{Song2019}%
  \BibitemOpen
  \bibfield  {author} {\bibinfo {author} {\bibfnamefont {T.}~\bibnamefont
  {Song}}, \bibinfo {author} {\bibfnamefont {M.~W.~Y.}\ \bibnamefont {Tu}},
  \bibinfo {author} {\bibfnamefont {C.}~\bibnamefont {Carnahan}}, \bibinfo
  {author} {\bibfnamefont {X.}~\bibnamefont {Cai}}, \bibinfo {author}
  {\bibfnamefont {T.}~\bibnamefont {Taniguchi}}, \bibinfo {author}
  {\bibfnamefont {K.}~\bibnamefont {Watanabe}}, \bibinfo {author}
  {\bibfnamefont {M.~A.}\ \bibnamefont {McGuire}}, \bibinfo {author}
  {\bibfnamefont {D.~H.}\ \bibnamefont {Cobden}}, \bibinfo {author}
  {\bibfnamefont {D.}~\bibnamefont {Xiao}}, \bibinfo {author} {\bibfnamefont
  {W.}~\bibnamefont {Yao}},\ and\ \bibinfo {author} {\bibfnamefont
  {X.}~\bibnamefont {Xu}},\ }\bibfield  {title} {\bibinfo {title} {{Voltage
  Control of a van der Waals Spin-Filter Magnetic Tunnel Junction}},\ }\href
  {https://doi.org/10.1021/acs.nanolett.8b04160} {\bibfield  {journal}
  {\bibinfo  {journal} {Nano Letters}\ }\textbf {\bibinfo {volume} {19}},\
  \bibinfo {pages} {915} (\bibinfo {year} {2019})}\BibitemShut {NoStop}%
\bibitem [{\citenamefont {Jiang}\ \emph {et~al.}(2019)\citenamefont {Jiang},
  \citenamefont {Li}, \citenamefont {Wang}, \citenamefont {Shan},\ and\
  \citenamefont {Mak}}]{Jiang2019}%
  \BibitemOpen
  \bibfield  {author} {\bibinfo {author} {\bibfnamefont {S.}~\bibnamefont
  {Jiang}}, \bibinfo {author} {\bibfnamefont {L.}~\bibnamefont {Li}}, \bibinfo
  {author} {\bibfnamefont {Z.}~\bibnamefont {Wang}}, \bibinfo {author}
  {\bibfnamefont {J.}~\bibnamefont {Shan}},\ and\ \bibinfo {author}
  {\bibfnamefont {K.~F.}\ \bibnamefont {Mak}},\ }\bibfield  {title} {\bibinfo
  {title} {{Spin tunnel field-effect transistors based on two-dimensional van
  der Waals heterostructures}},\ }\href
  {https://doi.org/10.1038/s41928-019-0232-3} {\bibfield  {journal} {\bibinfo
  {journal} {Nature Electronics}\ }\textbf {\bibinfo {volume} {2}},\ \bibinfo
  {pages} {159} (\bibinfo {year} {2019})}\BibitemShut {NoStop}%
\bibitem [{\citenamefont {Telford}\ \emph {et~al.}(2022)\citenamefont
  {Telford}, \citenamefont {Dismukes}, \citenamefont {Dudley}, \citenamefont
  {Wiscons}, \citenamefont {Lee}, \citenamefont {Chica}, \citenamefont
  {Ziebel}, \citenamefont {Han}, \citenamefont {Yu}, \citenamefont {Shabani},
  \citenamefont {Scheie}, \citenamefont {Watanabe}, \citenamefont {Taniguchi},
  \citenamefont {Xiao}, \citenamefont {Zhu}, \citenamefont {Pasupathy},
  \citenamefont {Nuckolls}, \citenamefont {Zhu}, \citenamefont {Dean},\ and\
  \citenamefont {Roy}}]{Telford2022}%
  \BibitemOpen
  \bibfield  {author} {\bibinfo {author} {\bibfnamefont {E.~J.}\ \bibnamefont
  {Telford}}, \bibinfo {author} {\bibfnamefont {A.~H.}\ \bibnamefont
  {Dismukes}}, \bibinfo {author} {\bibfnamefont {R.~L.}\ \bibnamefont
  {Dudley}}, \bibinfo {author} {\bibfnamefont {R.~A.}\ \bibnamefont {Wiscons}},
  \bibinfo {author} {\bibfnamefont {K.}~\bibnamefont {Lee}}, \bibinfo {author}
  {\bibfnamefont {D.~G.}\ \bibnamefont {Chica}}, \bibinfo {author}
  {\bibfnamefont {M.~E.}\ \bibnamefont {Ziebel}}, \bibinfo {author}
  {\bibfnamefont {M.-G.}\ \bibnamefont {Han}}, \bibinfo {author} {\bibfnamefont
  {J.}~\bibnamefont {Yu}}, \bibinfo {author} {\bibfnamefont {S.}~\bibnamefont
  {Shabani}}, \bibinfo {author} {\bibfnamefont {A.}~\bibnamefont {Scheie}},
  \bibinfo {author} {\bibfnamefont {K.}~\bibnamefont {Watanabe}}, \bibinfo
  {author} {\bibfnamefont {T.}~\bibnamefont {Taniguchi}}, \bibinfo {author}
  {\bibfnamefont {D.}~\bibnamefont {Xiao}}, \bibinfo {author} {\bibfnamefont
  {Y.}~\bibnamefont {Zhu}}, \bibinfo {author} {\bibfnamefont {A.~N.}\
  \bibnamefont {Pasupathy}}, \bibinfo {author} {\bibfnamefont {C.}~\bibnamefont
  {Nuckolls}}, \bibinfo {author} {\bibfnamefont {X.}~\bibnamefont {Zhu}},
  \bibinfo {author} {\bibfnamefont {C.~R.}\ \bibnamefont {Dean}},\ and\
  \bibinfo {author} {\bibfnamefont {X.}~\bibnamefont {Roy}},\ }\bibfield
  {title} {\bibinfo {title} {{Coupling between magnetic order and charge
  transport in a two-dimensional magnetic semiconductor}},\ }\href
  {https://doi.org/10.1038/s41563-022-01245-x} {\bibfield  {journal} {\bibinfo
  {journal} {Nature Materials}\ }\textbf {\bibinfo {volume} {21}},\ \bibinfo
  {pages} {754} (\bibinfo {year} {2022})}\BibitemShut {NoStop}%
\bibitem [{\citenamefont {Wu}\ \emph {et~al.}(2022)\citenamefont {Wu},
  \citenamefont {Guti{\'{e}}rrez‐Lezama}, \citenamefont {L{\'{o}}pez‐Paz},
  \citenamefont {Gibertini}, \citenamefont {Watanabe}, \citenamefont
  {Taniguchi}, \citenamefont {von Rohr}, \citenamefont {Ubrig},\ and\
  \citenamefont {Morpurgo}}]{Wu2022a}%
  \BibitemOpen
  \bibfield  {author} {\bibinfo {author} {\bibfnamefont {F.}~\bibnamefont
  {Wu}}, \bibinfo {author} {\bibfnamefont {I.}~\bibnamefont
  {Guti{\'{e}}rrez‐Lezama}}, \bibinfo {author} {\bibfnamefont {S.~A.}\
  \bibnamefont {L{\'{o}}pez‐Paz}}, \bibinfo {author} {\bibfnamefont
  {M.}~\bibnamefont {Gibertini}}, \bibinfo {author} {\bibfnamefont
  {K.}~\bibnamefont {Watanabe}}, \bibinfo {author} {\bibfnamefont
  {T.}~\bibnamefont {Taniguchi}}, \bibinfo {author} {\bibfnamefont {F.~O.}\
  \bibnamefont {von Rohr}}, \bibinfo {author} {\bibfnamefont {N.}~\bibnamefont
  {Ubrig}},\ and\ \bibinfo {author} {\bibfnamefont {A.~F.}\ \bibnamefont
  {Morpurgo}},\ }\bibfield  {title} {\bibinfo {title} {{Quasi‐1D Electronic
  Transport in a 2D Magnetic Semiconductor}},\ }\href
  {https://doi.org/10.1002/adma.202109759} {\bibfield  {journal} {\bibinfo
  {journal} {Advanced Materials}\ }\textbf {\bibinfo {volume} {34}},\ \bibinfo
  {pages} {2109759} (\bibinfo {year} {2022})}\BibitemShut {NoStop}%
\bibitem [{\citenamefont {Lin}\ \emph {et~al.}(2025)\citenamefont {Lin},
  \citenamefont {Wu}, \citenamefont {Ubrig}, \citenamefont {Liao},
  \citenamefont {Yao}, \citenamefont {Guti{\'{e}}rrez-Lezama},\ and\
  \citenamefont {Morpurgo}}]{Lin2024}%
  \BibitemOpen
  \bibfield  {author} {\bibinfo {author} {\bibfnamefont {X.}~\bibnamefont
  {Lin}}, \bibinfo {author} {\bibfnamefont {F.}~\bibnamefont {Wu}}, \bibinfo
  {author} {\bibfnamefont {N.}~\bibnamefont {Ubrig}}, \bibinfo {author}
  {\bibfnamefont {M.}~\bibnamefont {Liao}}, \bibinfo {author} {\bibfnamefont
  {F.}~\bibnamefont {Yao}}, \bibinfo {author} {\bibfnamefont {I.}~\bibnamefont
  {Guti{\'{e}}rrez-Lezama}},\ and\ \bibinfo {author} {\bibfnamefont {A.~F.}\
  \bibnamefont {Morpurgo}},\ }\bibfield  {title} {\bibinfo {title} {{Positive
  Oscillating Magnetoresistance in a van der Waals Antiferromagnetic
  Semiconductor}},\ }\href {https://doi.org/10.1103/PhysRevX.15.011017}
  {\bibfield  {journal} {\bibinfo  {journal} {Physical Review X}\ }\textbf
  {\bibinfo {volume} {15}},\ \bibinfo {pages} {011017} (\bibinfo {year}
  {2025})}\BibitemShut {NoStop}%
\bibitem [{\citenamefont {Mott}\ and\ \citenamefont {Gurney}(1948)}]{Mott1950}%
  \BibitemOpen
  \bibfield  {author} {\bibinfo {author} {\bibfnamefont {N.~F.}\ \bibnamefont
  {Mott}}\ and\ \bibinfo {author} {\bibfnamefont {R.~W.}\ \bibnamefont
  {Gurney}},\ }\href@noop {} {\emph {\bibinfo {title} {{Electronic processes in
  ionic crystals}}}},\ \bibinfo {edition} {2nd}\ ed.\ (\bibinfo  {publisher}
  {Oxford University Press},\ \bibinfo {address} {London},\ \bibinfo {year}
  {1948})\BibitemShut {NoStop}%
\bibitem [{\citenamefont {Rose}(1955)}]{Rose1955}%
  \BibitemOpen
  \bibfield  {author} {\bibinfo {author} {\bibfnamefont {A.}~\bibnamefont
  {Rose}},\ }\bibfield  {title} {\bibinfo {title} {{Space-Charge-Limited
  Currents in Solids}},\ }\href {https://doi.org/10.1103/PhysRev.97.1538}
  {\bibfield  {journal} {\bibinfo  {journal} {Physical Review}\ }\textbf
  {\bibinfo {volume} {97}},\ \bibinfo {pages} {1538} (\bibinfo {year}
  {1955})}\BibitemShut {NoStop}%
\bibitem [{\citenamefont {Mark}\ and\ \citenamefont
  {Helfrich}(1962)}]{Mark1962}%
  \BibitemOpen
  \bibfield  {author} {\bibinfo {author} {\bibfnamefont {P.}~\bibnamefont
  {Mark}}\ and\ \bibinfo {author} {\bibfnamefont {W.}~\bibnamefont
  {Helfrich}},\ }\bibfield  {title} {\bibinfo {title} {{Space-Charge-Limited
  Currents in Organic Crystals}},\ }\href {https://doi.org/10.1063/1.1728487}
  {\bibfield  {journal} {\bibinfo  {journal} {Journal of Applied Physics}\
  }\textbf {\bibinfo {volume} {33}},\ \bibinfo {pages} {205} (\bibinfo {year}
  {1962})}\BibitemShut {NoStop}%
\bibitem [{\citenamefont {Lampert}(1964)}]{Lampert1964}%
  \BibitemOpen
  \bibfield  {author} {\bibinfo {author} {\bibfnamefont {M.~A.}\ \bibnamefont
  {Lampert}},\ }\bibfield  {title} {\bibinfo {title} {{Volume-controlled
  current injection in insulators}},\ }\href
  {https://doi.org/10.1088/0034-4885/27/1/307} {\bibfield  {journal} {\bibinfo
  {journal} {Reports on Progress in Physics}\ }\textbf {\bibinfo {volume}
  {27}},\ \bibinfo {pages} {307} (\bibinfo {year} {1964})}\BibitemShut
  {NoStop}%
\bibitem [{\citenamefont {Shi}\ \emph {et~al.}(2015)\citenamefont {Shi},
  \citenamefont {Adinolfi}, \citenamefont {Comin}, \citenamefont {Yuan},
  \citenamefont {Alarousu}, \citenamefont {Buin}, \citenamefont {Chen},
  \citenamefont {Hoogland}, \citenamefont {Rothenberger}, \citenamefont
  {Katsiev}, \citenamefont {Losovyj}, \citenamefont {Zhang}, \citenamefont
  {Dowben}, \citenamefont {Mohammed}, \citenamefont {Sargent},\ and\
  \citenamefont {Bakr}}]{Shi2015a}%
  \BibitemOpen
  \bibfield  {author} {\bibinfo {author} {\bibfnamefont {D.}~\bibnamefont
  {Shi}}, \bibinfo {author} {\bibfnamefont {V.}~\bibnamefont {Adinolfi}},
  \bibinfo {author} {\bibfnamefont {R.}~\bibnamefont {Comin}}, \bibinfo
  {author} {\bibfnamefont {M.}~\bibnamefont {Yuan}}, \bibinfo {author}
  {\bibfnamefont {E.}~\bibnamefont {Alarousu}}, \bibinfo {author}
  {\bibfnamefont {A.}~\bibnamefont {Buin}}, \bibinfo {author} {\bibfnamefont
  {Y.}~\bibnamefont {Chen}}, \bibinfo {author} {\bibfnamefont {S.}~\bibnamefont
  {Hoogland}}, \bibinfo {author} {\bibfnamefont {A.}~\bibnamefont
  {Rothenberger}}, \bibinfo {author} {\bibfnamefont {K.}~\bibnamefont
  {Katsiev}}, \bibinfo {author} {\bibfnamefont {Y.}~\bibnamefont {Losovyj}},
  \bibinfo {author} {\bibfnamefont {X.}~\bibnamefont {Zhang}}, \bibinfo
  {author} {\bibfnamefont {P.~A.}\ \bibnamefont {Dowben}}, \bibinfo {author}
  {\bibfnamefont {O.~F.}\ \bibnamefont {Mohammed}}, \bibinfo {author}
  {\bibfnamefont {E.~H.}\ \bibnamefont {Sargent}},\ and\ \bibinfo {author}
  {\bibfnamefont {O.~M.}\ \bibnamefont {Bakr}},\ }\bibfield  {title} {\bibinfo
  {title} {{Low trap-state density and long carrier diffusion in organolead
  trihalide perovskite single crystals}},\ }\href
  {https://doi.org/10.1126/science.aaa2725} {\bibfield  {journal} {\bibinfo
  {journal} {Science}\ }\textbf {\bibinfo {volume} {347}},\ \bibinfo {pages}
  {519} (\bibinfo {year} {2015})}\BibitemShut {NoStop}%
\bibitem [{\citenamefont {Jedema}\ \emph {et~al.}(2001)\citenamefont {Jedema},
  \citenamefont {Filip},\ and\ \citenamefont {van Wees}}]{Jedema2001}%
  \BibitemOpen
  \bibfield  {author} {\bibinfo {author} {\bibfnamefont {F.~J.}\ \bibnamefont
  {Jedema}}, \bibinfo {author} {\bibfnamefont {A.~T.}\ \bibnamefont {Filip}},\
  and\ \bibinfo {author} {\bibfnamefont {B.~J.}\ \bibnamefont {van Wees}},\
  }\bibfield  {title} {\bibinfo {title} {{Electrical spin injection and
  accumulation at room temperature in an all-metal mesoscopic spin valve}},\
  }\href {https://doi.org/10.1038/35066533} {\bibfield  {journal} {\bibinfo
  {journal} {Nature}\ }\textbf {\bibinfo {volume} {410}},\ \bibinfo {pages}
  {345} (\bibinfo {year} {2001})}\BibitemShut {NoStop}%
\bibitem [{\citenamefont {Thiel}\ \emph {et~al.}(2019)\citenamefont {Thiel},
  \citenamefont {Wang}, \citenamefont {Tschudin}, \citenamefont {Rohner},
  \citenamefont {Guti{\'{e}}rrez-Lezama}, \citenamefont {Ubrig}, \citenamefont
  {Gibertini}, \citenamefont {Giannini}, \citenamefont {Morpurgo},\ and\
  \citenamefont {Maletinsky}}]{Thiel2019}%
  \BibitemOpen
  \bibfield  {author} {\bibinfo {author} {\bibfnamefont {L.}~\bibnamefont
  {Thiel}}, \bibinfo {author} {\bibfnamefont {Z.}~\bibnamefont {Wang}},
  \bibinfo {author} {\bibfnamefont {M.~A.}\ \bibnamefont {Tschudin}}, \bibinfo
  {author} {\bibfnamefont {D.}~\bibnamefont {Rohner}}, \bibinfo {author}
  {\bibfnamefont {I.}~\bibnamefont {Guti{\'{e}}rrez-Lezama}}, \bibinfo {author}
  {\bibfnamefont {N.}~\bibnamefont {Ubrig}}, \bibinfo {author} {\bibfnamefont
  {M.}~\bibnamefont {Gibertini}}, \bibinfo {author} {\bibfnamefont
  {E.}~\bibnamefont {Giannini}}, \bibinfo {author} {\bibfnamefont {A.~F.}\
  \bibnamefont {Morpurgo}},\ and\ \bibinfo {author} {\bibfnamefont
  {P.}~\bibnamefont {Maletinsky}},\ }\bibfield  {title} {\bibinfo {title}
  {{Probing magnetism in 2D materials at the nanoscale with single-spin
  microscopy}},\ }\href {https://doi.org/10.1126/science.aav6926} {\bibfield
  {journal} {\bibinfo  {journal} {Science}\ }\textbf {\bibinfo {volume}
  {364}},\ \bibinfo {pages} {973} (\bibinfo {year} {2019})}\BibitemShut
  {NoStop}%
\bibitem [{\citenamefont {Song}\ \emph {et~al.}(2021)\citenamefont {Song},
  \citenamefont {Sun}, \citenamefont {Anderson}, \citenamefont {Wang},
  \citenamefont {Qian}, \citenamefont {Taniguchi}, \citenamefont {Watanabe},
  \citenamefont {McGuire}, \citenamefont {St{\"{o}}hr}, \citenamefont {Xiao},
  \citenamefont {Cao}, \citenamefont {Wrachtrup},\ and\ \citenamefont
  {Xu}}]{Song2021}%
  \BibitemOpen
  \bibfield  {author} {\bibinfo {author} {\bibfnamefont {T.}~\bibnamefont
  {Song}}, \bibinfo {author} {\bibfnamefont {Q.-C.}\ \bibnamefont {Sun}},
  \bibinfo {author} {\bibfnamefont {E.}~\bibnamefont {Anderson}}, \bibinfo
  {author} {\bibfnamefont {C.}~\bibnamefont {Wang}}, \bibinfo {author}
  {\bibfnamefont {J.}~\bibnamefont {Qian}}, \bibinfo {author} {\bibfnamefont
  {T.}~\bibnamefont {Taniguchi}}, \bibinfo {author} {\bibfnamefont
  {K.}~\bibnamefont {Watanabe}}, \bibinfo {author} {\bibfnamefont {M.~A.}\
  \bibnamefont {McGuire}}, \bibinfo {author} {\bibfnamefont {R.}~\bibnamefont
  {St{\"{o}}hr}}, \bibinfo {author} {\bibfnamefont {D.}~\bibnamefont {Xiao}},
  \bibinfo {author} {\bibfnamefont {T.}~\bibnamefont {Cao}}, \bibinfo {author}
  {\bibfnamefont {J.}~\bibnamefont {Wrachtrup}},\ and\ \bibinfo {author}
  {\bibfnamefont {X.}~\bibnamefont {Xu}},\ }\bibfield  {title} {\bibinfo
  {title} {{Direct visualization of magnetic domains and moir{\'{e}} magnetism
  in twisted 2D magnets}},\ }\href {https://doi.org/10.1126/science.abj7478}
  {\bibfield  {journal} {\bibinfo  {journal} {Science}\ }\textbf {\bibinfo
  {volume} {374}},\ \bibinfo {pages} {1140} (\bibinfo {year}
  {2021})}\BibitemShut {NoStop}%
\bibitem [{\citenamefont {Wang}\ \emph {et~al.}(2025)\citenamefont {Wang},
  \citenamefont {Graham}, \citenamefont {Rama-Eiroa}, \citenamefont {Islam},
  \citenamefont {Badarneh}, \citenamefont {{Nunes Gontijo}}, \citenamefont
  {Tiwari}, \citenamefont {Adhikari}, \citenamefont {Zhang}, \citenamefont
  {Watanabe}, \citenamefont {Taniguchi}, \citenamefont {Besson}, \citenamefont
  {{J. G. Santos}}, \citenamefont {Lin},\ and\ \citenamefont
  {Zhou}}]{Wang2025a}%
  \BibitemOpen
  \bibfield  {author} {\bibinfo {author} {\bibfnamefont {Y.-X.}\ \bibnamefont
  {Wang}}, \bibinfo {author} {\bibfnamefont {T.~K.~M.}\ \bibnamefont {Graham}},
  \bibinfo {author} {\bibfnamefont {R.}~\bibnamefont {Rama-Eiroa}}, \bibinfo
  {author} {\bibfnamefont {M.~A.}\ \bibnamefont {Islam}}, \bibinfo {author}
  {\bibfnamefont {M.~H.}\ \bibnamefont {Badarneh}}, \bibinfo {author}
  {\bibfnamefont {R.}~\bibnamefont {{Nunes Gontijo}}}, \bibinfo {author}
  {\bibfnamefont {G.~P.}\ \bibnamefont {Tiwari}}, \bibinfo {author}
  {\bibfnamefont {T.}~\bibnamefont {Adhikari}}, \bibinfo {author}
  {\bibfnamefont {X.-Y.}\ \bibnamefont {Zhang}}, \bibinfo {author}
  {\bibfnamefont {K.}~\bibnamefont {Watanabe}}, \bibinfo {author}
  {\bibfnamefont {T.}~\bibnamefont {Taniguchi}}, \bibinfo {author}
  {\bibfnamefont {C.}~\bibnamefont {Besson}}, \bibinfo {author} {\bibfnamefont
  {E.}~\bibnamefont {{J. G. Santos}}}, \bibinfo {author} {\bibfnamefont
  {Z.}~\bibnamefont {Lin}},\ and\ \bibinfo {author} {\bibfnamefont {B.~B.}\
  \bibnamefont {Zhou}},\ }\bibfield  {title} {\bibinfo {title} {{Configurable
  antiferromagnetic domains and lateral exchange bias in atomically thin
  CrPS4}},\ }\href {https://doi.org/10.1038/s41563-025-02259-x} {\bibfield
  {journal} {\bibinfo  {journal} {Nature Materials}\ }\textbf {\bibinfo
  {volume} {24}},\ \bibinfo {pages} {1414} (\bibinfo {year}
  {2025})}\BibitemShut {NoStop}%
\bibitem [{\citenamefont {Pellet-Mary}\ \emph {et~al.}(2025)\citenamefont
  {Pellet-Mary}, \citenamefont {Dutta}, \citenamefont {Tschudin}, \citenamefont
  {Siegwolf}, \citenamefont {Gross}, \citenamefont {Broadway}, \citenamefont
  {Cox}, \citenamefont {Schrader}, \citenamefont {Happacher}, \citenamefont
  {Chica}, \citenamefont {Dean}, \citenamefont {Roy},\ and\ \citenamefont
  {Maletinsky}}]{Pellet-Mary2025}%
  \BibitemOpen
  \bibfield  {author} {\bibinfo {author} {\bibfnamefont {C.}~\bibnamefont
  {Pellet-Mary}}, \bibinfo {author} {\bibfnamefont {D.}~\bibnamefont {Dutta}},
  \bibinfo {author} {\bibfnamefont {M.~A.}\ \bibnamefont {Tschudin}}, \bibinfo
  {author} {\bibfnamefont {P.}~\bibnamefont {Siegwolf}}, \bibinfo {author}
  {\bibfnamefont {B.}~\bibnamefont {Gross}}, \bibinfo {author} {\bibfnamefont
  {D.~A.}\ \bibnamefont {Broadway}}, \bibinfo {author} {\bibfnamefont
  {J.}~\bibnamefont {Cox}}, \bibinfo {author} {\bibfnamefont {C.}~\bibnamefont
  {Schrader}}, \bibinfo {author} {\bibfnamefont {J.}~\bibnamefont {Happacher}},
  \bibinfo {author} {\bibfnamefont {D.~G.}\ \bibnamefont {Chica}}, \bibinfo
  {author} {\bibfnamefont {C.~R.}\ \bibnamefont {Dean}}, \bibinfo {author}
  {\bibfnamefont {X.}~\bibnamefont {Roy}},\ and\ \bibinfo {author}
  {\bibfnamefont {P.}~\bibnamefont {Maletinsky}},\ }\bibfield  {title}
  {\bibinfo {title} {{Lateral exchange bias for N{\'{e}}el-vector control in
  atomically thin antiferromagnets}},\ }\href
  {https://doi.org/10.1038/s41467-025-64700-8} {\bibfield  {journal} {\bibinfo
  {journal} {Nature Communications}\ }\textbf {\bibinfo {volume} {16}},\
  \bibinfo {pages} {9725} (\bibinfo {year} {2025})}\BibitemShut {NoStop}%
\bibitem [{\citenamefont {Weile}\ \emph {et~al.}(2025)\citenamefont {Weile},
  \citenamefont {Grytsiuk}, \citenamefont {Penn}, \citenamefont {Chica},
  \citenamefont {Roy}, \citenamefont {Mosina}, \citenamefont {Sofer},
  \citenamefont {Schi{\o}tz}, \citenamefont {Helveg}, \citenamefont
  {R{\"{o}}sner}, \citenamefont {Ross},\ and\ \citenamefont
  {Klein}}]{Weile2025}%
  \BibitemOpen
  \bibfield  {author} {\bibinfo {author} {\bibfnamefont {M.}~\bibnamefont
  {Weile}}, \bibinfo {author} {\bibfnamefont {S.}~\bibnamefont {Grytsiuk}},
  \bibinfo {author} {\bibfnamefont {A.}~\bibnamefont {Penn}}, \bibinfo {author}
  {\bibfnamefont {D.~G.}\ \bibnamefont {Chica}}, \bibinfo {author}
  {\bibfnamefont {X.}~\bibnamefont {Roy}}, \bibinfo {author} {\bibfnamefont
  {K.}~\bibnamefont {Mosina}}, \bibinfo {author} {\bibfnamefont
  {Z.}~\bibnamefont {Sofer}}, \bibinfo {author} {\bibfnamefont
  {J.}~\bibnamefont {Schi{\o}tz}}, \bibinfo {author} {\bibfnamefont
  {S.}~\bibnamefont {Helveg}}, \bibinfo {author} {\bibfnamefont
  {M.}~\bibnamefont {R{\"{o}}sner}}, \bibinfo {author} {\bibfnamefont {F.~M.}\
  \bibnamefont {Ross}},\ and\ \bibinfo {author} {\bibfnamefont
  {J.}~\bibnamefont {Klein}},\ }\bibfield  {title} {\bibinfo {title} {{Defect
  Complexes in CrSBr Revealed Through Electron Microscopy and Deep Learning}},\
  }\href {https://doi.org/10.1103/PhysRevX.15.021080} {\bibfield  {journal}
  {\bibinfo  {journal} {Physical Review X}\ }\textbf {\bibinfo {volume} {15}},\
  \bibinfo {pages} {021080} (\bibinfo {year} {2025})}\BibitemShut {NoStop}%
\bibitem [{\citenamefont {Boix‐Constant}\ \emph {et~al.}(2022)\citenamefont
  {Boix‐Constant}, \citenamefont {Ma{\~{n}}as‐Valero}, \citenamefont
  {Ruiz}, \citenamefont {Rybakov}, \citenamefont {Konieczny}, \citenamefont
  {Pillet}, \citenamefont {Baldov{\'{i}}},\ and\ \citenamefont
  {Coronado}}]{Boix-Constant2022}%
  \BibitemOpen
  \bibfield  {author} {\bibinfo {author} {\bibfnamefont {C.}~\bibnamefont
  {Boix‐Constant}}, \bibinfo {author} {\bibfnamefont {S.}~\bibnamefont
  {Ma{\~{n}}as‐Valero}}, \bibinfo {author} {\bibfnamefont {A.~M.}\
  \bibnamefont {Ruiz}}, \bibinfo {author} {\bibfnamefont {A.}~\bibnamefont
  {Rybakov}}, \bibinfo {author} {\bibfnamefont {K.~A.}\ \bibnamefont
  {Konieczny}}, \bibinfo {author} {\bibfnamefont {S.}~\bibnamefont {Pillet}},
  \bibinfo {author} {\bibfnamefont {J.~J.}\ \bibnamefont {Baldov{\'{i}}}},\
  and\ \bibinfo {author} {\bibfnamefont {E.}~\bibnamefont {Coronado}},\
  }\bibfield  {title} {\bibinfo {title} {{Probing the Spin Dimensionality in
  Single‐Layer CrSBr Van Der Waals Heterostructures by Magneto‐Transport
  Measurements}},\ }\href {https://doi.org/10.1002/adma.202204940} {\bibfield
  {journal} {\bibinfo  {journal} {Advanced Materials}\ }\textbf {\bibinfo
  {volume} {34}},\ \bibinfo {pages} {2204940} (\bibinfo {year}
  {2022})}\BibitemShut {NoStop}%
\bibitem [{Note1()}]{Note1}%
  \BibitemOpen
  \bibinfo {note} {For the basic theory of SCL currents, details on the
  experimental setup and data analysis, and additional supporting data, see
  Supplemental Material at [url], which includes Refs. \cite
  {Gross2017,Dirnberger2023,Wang2017d}. Additional data include temperature and
  magnetic field angle dependence, alternative field training conditions,
  extended gate voltages, and a second SCST device.}\BibitemShut {Stop}%
\bibitem [{\citenamefont {Liu}\ \emph {et~al.}(2022)\citenamefont {Liu},
  \citenamefont {Choi}, \citenamefont {Hwang}, \citenamefont {Yoo},\ and\
  \citenamefont {Sun}}]{Liu2022b}%
  \BibitemOpen
  \bibfield  {author} {\bibinfo {author} {\bibfnamefont {X.}~\bibnamefont
  {Liu}}, \bibinfo {author} {\bibfnamefont {M.~S.}\ \bibnamefont {Choi}},
  \bibinfo {author} {\bibfnamefont {E.}~\bibnamefont {Hwang}}, \bibinfo
  {author} {\bibfnamefont {W.~J.}\ \bibnamefont {Yoo}},\ and\ \bibinfo {author}
  {\bibfnamefont {J.}~\bibnamefont {Sun}},\ }\bibfield  {title} {\bibinfo
  {title} {{Fermi Level Pinning Dependent 2D Semiconductor Devices: Challenges
  and Prospects}},\ }\href {https://doi.org/10.1002/adma.202108425} {\bibfield
  {journal} {\bibinfo  {journal} {Advanced Materials}\ }\textbf {\bibinfo
  {volume} {34}},\ \bibinfo {pages} {2108425} (\bibinfo {year}
  {2022})}\BibitemShut {NoStop}%
\bibitem [{\citenamefont {Geurst}(1966)}]{Geurst1966}%
  \BibitemOpen
  \bibfield  {author} {\bibinfo {author} {\bibfnamefont {J.~A.}\ \bibnamefont
  {Geurst}},\ }\bibfield  {title} {\bibinfo {title} {{Theory of
  Space-Charge-Limited Currents in Thin Semiconductor Layers}},\ }\href
  {https://doi.org/10.1515/9783112492505-009} {\bibfield  {journal} {\bibinfo
  {journal} {Phys. Status Solidi}\ }\textbf {\bibinfo {volume} {15}},\ \bibinfo
  {pages} {107} (\bibinfo {year} {1966})}\BibitemShut {NoStop}%
\bibitem [{\citenamefont {Lin}\ \emph {et~al.}(2024)\citenamefont {Lin},
  \citenamefont {Wu}, \citenamefont {L{\'{o}}pez-Paz}, \citenamefont {von
  Rohr}, \citenamefont {Gibertini}, \citenamefont {Guti{\'{e}}rrez-Lezama},\
  and\ \citenamefont {Morpurgo}}]{Lin2024a}%
  \BibitemOpen
  \bibfield  {author} {\bibinfo {author} {\bibfnamefont {X.}~\bibnamefont
  {Lin}}, \bibinfo {author} {\bibfnamefont {F.}~\bibnamefont {Wu}}, \bibinfo
  {author} {\bibfnamefont {S.~A.}\ \bibnamefont {L{\'{o}}pez-Paz}}, \bibinfo
  {author} {\bibfnamefont {F.~O.}\ \bibnamefont {von Rohr}}, \bibinfo {author}
  {\bibfnamefont {M.}~\bibnamefont {Gibertini}}, \bibinfo {author}
  {\bibfnamefont {I.}~\bibnamefont {Guti{\'{e}}rrez-Lezama}},\ and\ \bibinfo
  {author} {\bibfnamefont {A.~F.}\ \bibnamefont {Morpurgo}},\ }\bibfield
  {title} {\bibinfo {title} {{Influence of magnetism on vertical hopping
  transport in CrSBr}},\ }\href
  {https://doi.org/10.1103/PhysRevResearch.6.013185} {\bibfield  {journal}
  {\bibinfo  {journal} {Physical Review Research}\ }\textbf {\bibinfo {volume}
  {6}},\ \bibinfo {pages} {013185} (\bibinfo {year} {2024})}\BibitemShut
  {NoStop}%
\bibitem [{\citenamefont {Rizzo}\ \emph {et~al.}(2025)\citenamefont {Rizzo},
  \citenamefont {Seewald}, \citenamefont {Zhao}, \citenamefont {Cox},
  \citenamefont {Xie}, \citenamefont {Vitalone}, \citenamefont {Ruta},
  \citenamefont {Chica}, \citenamefont {Shao}, \citenamefont {Shabani},
  \citenamefont {Telford}, \citenamefont {Strasbourg}, \citenamefont
  {Darlington}, \citenamefont {Xu}, \citenamefont {Qiu}, \citenamefont
  {Devarakonda}, \citenamefont {Taniguchi}, \citenamefont {Watanabe},
  \citenamefont {Zhu}, \citenamefont {Schuck}, \citenamefont {Dean},
  \citenamefont {Roy}, \citenamefont {Millis}, \citenamefont {Cao},
  \citenamefont {Rubio}, \citenamefont {Pasupathy},\ and\ \citenamefont
  {Basov}}]{Rizzo2025}%
  \BibitemOpen
  \bibfield  {author} {\bibinfo {author} {\bibfnamefont {D.~J.}\ \bibnamefont
  {Rizzo}}, \bibinfo {author} {\bibfnamefont {E.}~\bibnamefont {Seewald}},
  \bibinfo {author} {\bibfnamefont {F.}~\bibnamefont {Zhao}}, \bibinfo {author}
  {\bibfnamefont {J.}~\bibnamefont {Cox}}, \bibinfo {author} {\bibfnamefont
  {K.}~\bibnamefont {Xie}}, \bibinfo {author} {\bibfnamefont {R.~A.}\
  \bibnamefont {Vitalone}}, \bibinfo {author} {\bibfnamefont {F.~L.}\
  \bibnamefont {Ruta}}, \bibinfo {author} {\bibfnamefont {D.~G.}\ \bibnamefont
  {Chica}}, \bibinfo {author} {\bibfnamefont {Y.}~\bibnamefont {Shao}},
  \bibinfo {author} {\bibfnamefont {S.}~\bibnamefont {Shabani}}, \bibinfo
  {author} {\bibfnamefont {E.~J.}\ \bibnamefont {Telford}}, \bibinfo {author}
  {\bibfnamefont {M.~C.}\ \bibnamefont {Strasbourg}}, \bibinfo {author}
  {\bibfnamefont {T.~P.}\ \bibnamefont {Darlington}}, \bibinfo {author}
  {\bibfnamefont {S.}~\bibnamefont {Xu}}, \bibinfo {author} {\bibfnamefont
  {S.}~\bibnamefont {Qiu}}, \bibinfo {author} {\bibfnamefont {A.}~\bibnamefont
  {Devarakonda}}, \bibinfo {author} {\bibfnamefont {T.}~\bibnamefont
  {Taniguchi}}, \bibinfo {author} {\bibfnamefont {K.}~\bibnamefont {Watanabe}},
  \bibinfo {author} {\bibfnamefont {X.}~\bibnamefont {Zhu}}, \bibinfo {author}
  {\bibfnamefont {P.~J.}\ \bibnamefont {Schuck}}, \bibinfo {author}
  {\bibfnamefont {C.~R.}\ \bibnamefont {Dean}}, \bibinfo {author}
  {\bibfnamefont {X.}~\bibnamefont {Roy}}, \bibinfo {author} {\bibfnamefont
  {A.~J.}\ \bibnamefont {Millis}}, \bibinfo {author} {\bibfnamefont
  {T.}~\bibnamefont {Cao}}, \bibinfo {author} {\bibfnamefont {A.}~\bibnamefont
  {Rubio}}, \bibinfo {author} {\bibfnamefont {A.~N.}\ \bibnamefont
  {Pasupathy}},\ and\ \bibinfo {author} {\bibfnamefont {D.~N.}\ \bibnamefont
  {Basov}},\ }\bibfield  {title} {\bibinfo {title} {{Engineering anisotropic
  electrodynamics at the graphene/CrSBr interface}},\ }\href
  {https://doi.org/10.1038/s41467-025-56804-y} {\bibfield  {journal} {\bibinfo
  {journal} {Nature Communications}\ }\textbf {\bibinfo {volume} {16}},\
  \bibinfo {pages} {1853} (\bibinfo {year} {2025})}\BibitemShut {NoStop}%
\bibitem [{\citenamefont {Chou}\ \emph {et~al.}(2025)\citenamefont {Chou},
  \citenamefont {Park}, \citenamefont {Ingla-Aynes}, \citenamefont {Klein},
  \citenamefont {Mosina}, \citenamefont {Moodera}, \citenamefont {Sofer},
  \citenamefont {Ross},\ and\ \citenamefont {Liu}}]{Chou2025}%
  \BibitemOpen
  \bibfield  {author} {\bibinfo {author} {\bibfnamefont {C.-T.}\ \bibnamefont
  {Chou}}, \bibinfo {author} {\bibfnamefont {E.}~\bibnamefont {Park}}, \bibinfo
  {author} {\bibfnamefont {J.}~\bibnamefont {Ingla-Aynes}}, \bibinfo {author}
  {\bibfnamefont {J.}~\bibnamefont {Klein}}, \bibinfo {author} {\bibfnamefont
  {K.}~\bibnamefont {Mosina}}, \bibinfo {author} {\bibfnamefont {J.~S.}\
  \bibnamefont {Moodera}}, \bibinfo {author} {\bibfnamefont {Z.}~\bibnamefont
  {Sofer}}, \bibinfo {author} {\bibfnamefont {F.~M.}\ \bibnamefont {Ross}},\
  and\ \bibinfo {author} {\bibfnamefont {L.}~\bibnamefont {Liu}},\ }\bibfield
  {title} {\bibinfo {title} {{Large Magnetoresistance in an Electrically
  Tunable van der Waals Antiferromagnet}},\ }\href
  {https://doi.org/10.1103/hpmq-rnh4} {\bibfield  {journal} {\bibinfo
  {journal} {Physical Review Letters}\ }\textbf {\bibinfo {volume} {135}},\
  \bibinfo {pages} {136702} (\bibinfo {year} {2025})}\BibitemShut {NoStop}%
\bibitem [{\citenamefont {Guo}\ \emph {et~al.}(2018)\citenamefont {Guo},
  \citenamefont {Zhang}, \citenamefont {Yuan}, \citenamefont {Wang},\ and\
  \citenamefont {Wang}}]{Guo2018c}%
  \BibitemOpen
  \bibfield  {author} {\bibinfo {author} {\bibfnamefont {Y.}~\bibnamefont
  {Guo}}, \bibinfo {author} {\bibfnamefont {Y.}~\bibnamefont {Zhang}}, \bibinfo
  {author} {\bibfnamefont {S.}~\bibnamefont {Yuan}}, \bibinfo {author}
  {\bibfnamefont {B.}~\bibnamefont {Wang}},\ and\ \bibinfo {author}
  {\bibfnamefont {J.}~\bibnamefont {Wang}},\ }\bibfield  {title} {\bibinfo
  {title} {{Chromium sulfide halide monolayers: intrinsic ferromagnetic
  semiconductors with large spin polarization and high carrier mobility}},\
  }\href {https://doi.org/10.1039/C8NR06368K} {\bibfield  {journal} {\bibinfo
  {journal} {Nanoscale}\ }\textbf {\bibinfo {volume} {10}},\ \bibinfo {pages}
  {18036} (\bibinfo {year} {2018})}\BibitemShut {NoStop}%
\bibitem [{\citenamefont {Shklovskii}\ and\ \citenamefont
  {Efros}(1984)}]{Shklovskii1984}%
  \BibitemOpen
  \bibfield  {author} {\bibinfo {author} {\bibfnamefont {B.~I.}\ \bibnamefont
  {Shklovskii}}\ and\ \bibinfo {author} {\bibfnamefont {A.~L.}\ \bibnamefont
  {Efros}},\ }\bibinfo {title} {{Variable-Range Hopping Conduction}},\ in\
  \href {https://doi.org/10.1007/978-3-662-02403-4_9} {\emph {\bibinfo
  {booktitle} {Electronic Properties of Doped Semiconductors}}}\ (\bibinfo
  {publisher} {Springer Berlin Heidelberg},\ \bibinfo {address} {Berlin,
  Heidelberg},\ \bibinfo {year} {1984})\ pp.\ \bibinfo {pages}
  {202--227}\BibitemShut {NoStop}%
\bibitem [{\citenamefont {Huang}\ \emph {et~al.}(2018)\citenamefont {Huang},
  \citenamefont {Clark}, \citenamefont {Klein}, \citenamefont {MacNeill},
  \citenamefont {Navarro-Moratalla}, \citenamefont {Seyler}, \citenamefont
  {Wilson}, \citenamefont {McGuire}, \citenamefont {Cobden}, \citenamefont
  {Xiao}, \citenamefont {Yao}, \citenamefont {Jarillo-Herrero},\ and\
  \citenamefont {Xu}}]{Huang2018}%
  \BibitemOpen
  \bibfield  {author} {\bibinfo {author} {\bibfnamefont {B.}~\bibnamefont
  {Huang}}, \bibinfo {author} {\bibfnamefont {G.}~\bibnamefont {Clark}},
  \bibinfo {author} {\bibfnamefont {D.~R.}\ \bibnamefont {Klein}}, \bibinfo
  {author} {\bibfnamefont {D.}~\bibnamefont {MacNeill}}, \bibinfo {author}
  {\bibfnamefont {E.}~\bibnamefont {Navarro-Moratalla}}, \bibinfo {author}
  {\bibfnamefont {K.~L.}\ \bibnamefont {Seyler}}, \bibinfo {author}
  {\bibfnamefont {N.}~\bibnamefont {Wilson}}, \bibinfo {author} {\bibfnamefont
  {M.~A.}\ \bibnamefont {McGuire}}, \bibinfo {author} {\bibfnamefont {D.~H.}\
  \bibnamefont {Cobden}}, \bibinfo {author} {\bibfnamefont {D.}~\bibnamefont
  {Xiao}}, \bibinfo {author} {\bibfnamefont {W.}~\bibnamefont {Yao}}, \bibinfo
  {author} {\bibfnamefont {P.}~\bibnamefont {Jarillo-Herrero}},\ and\ \bibinfo
  {author} {\bibfnamefont {X.}~\bibnamefont {Xu}},\ }\bibfield  {title}
  {\bibinfo {title} {{Electrical control of 2D magnetism in bilayer CrI3}},\
  }\href {https://doi.org/10.1038/s41565-018-0121-3} {\bibfield  {journal}
  {\bibinfo  {journal} {Nature Nanotechnology}\ }\textbf {\bibinfo {volume}
  {13}},\ \bibinfo {pages} {544} (\bibinfo {year} {2018})}\BibitemShut
  {NoStop}%
\bibitem [{\citenamefont {Jiang}\ \emph {et~al.}(2018)\citenamefont {Jiang},
  \citenamefont {Li}, \citenamefont {Wang}, \citenamefont {Mak},\ and\
  \citenamefont {Shan}}]{Jiang2018a}%
  \BibitemOpen
  \bibfield  {author} {\bibinfo {author} {\bibfnamefont {S.}~\bibnamefont
  {Jiang}}, \bibinfo {author} {\bibfnamefont {L.}~\bibnamefont {Li}}, \bibinfo
  {author} {\bibfnamefont {Z.}~\bibnamefont {Wang}}, \bibinfo {author}
  {\bibfnamefont {K.~F.}\ \bibnamefont {Mak}},\ and\ \bibinfo {author}
  {\bibfnamefont {J.}~\bibnamefont {Shan}},\ }\bibfield  {title} {\bibinfo
  {title} {{Controlling magnetism in 2D CrI3 by electrostatic doping}},\ }\href
  {https://doi.org/10.1038/s41565-018-0135-x} {\bibfield  {journal} {\bibinfo
  {journal} {Nature Nanotechnology}\ }\textbf {\bibinfo {volume} {13}},\
  \bibinfo {pages} {549} (\bibinfo {year} {2018})}\BibitemShut {NoStop}%
\bibitem [{\citenamefont {Tabataba-Vakili}\ \emph {et~al.}(2024)\citenamefont
  {Tabataba-Vakili}, \citenamefont {Nguyen}, \citenamefont {Rupp},
  \citenamefont {Mosina}, \citenamefont {Papavasileiou}, \citenamefont
  {Watanabe}, \citenamefont {Taniguchi}, \citenamefont {Maletinsky},
  \citenamefont {Glazov}, \citenamefont {Sofer}, \citenamefont {Baimuratov},\
  and\ \citenamefont {H{\"{o}}gele}}]{Tabataba-vakili2024}%
  \BibitemOpen
  \bibfield  {author} {\bibinfo {author} {\bibfnamefont {F.}~\bibnamefont
  {Tabataba-Vakili}}, \bibinfo {author} {\bibfnamefont {H.~P.~G.}\ \bibnamefont
  {Nguyen}}, \bibinfo {author} {\bibfnamefont {A.}~\bibnamefont {Rupp}},
  \bibinfo {author} {\bibfnamefont {K.}~\bibnamefont {Mosina}}, \bibinfo
  {author} {\bibfnamefont {A.}~\bibnamefont {Papavasileiou}}, \bibinfo {author}
  {\bibfnamefont {K.}~\bibnamefont {Watanabe}}, \bibinfo {author}
  {\bibfnamefont {T.}~\bibnamefont {Taniguchi}}, \bibinfo {author}
  {\bibfnamefont {P.}~\bibnamefont {Maletinsky}}, \bibinfo {author}
  {\bibfnamefont {M.~M.}\ \bibnamefont {Glazov}}, \bibinfo {author}
  {\bibfnamefont {Z.}~\bibnamefont {Sofer}}, \bibinfo {author} {\bibfnamefont
  {A.~S.}\ \bibnamefont {Baimuratov}},\ and\ \bibinfo {author} {\bibfnamefont
  {A.}~\bibnamefont {H{\"{o}}gele}},\ }\bibfield  {title} {\bibinfo {title}
  {{Doping-control of excitons and magnetism in few-layer CrSBr}},\ }\href
  {https://doi.org/10.1038/s41467-024-49048-9} {\bibfield  {journal} {\bibinfo
  {journal} {Nature Communications}\ }\textbf {\bibinfo {volume} {15}},\
  \bibinfo {pages} {4735} (\bibinfo {year} {2024})}\BibitemShut {NoStop}%
\bibitem [{\citenamefont {Sun}\ \emph {et~al.}(2025)\citenamefont {Sun},
  \citenamefont {Hong}, \citenamefont {Chen}, \citenamefont {Sheng},
  \citenamefont {Wu}, \citenamefont {Wang}, \citenamefont {Liang},
  \citenamefont {Liu}, \citenamefont {Yuan}, \citenamefont {Wu}, \citenamefont
  {Mi}, \citenamefont {Liu}, \citenamefont {Shen},\ and\ \citenamefont
  {Wu}}]{Sun2024}%
  \BibitemOpen
  \bibfield  {author} {\bibinfo {author} {\bibfnamefont {Z.}~\bibnamefont
  {Sun}}, \bibinfo {author} {\bibfnamefont {C.}~\bibnamefont {Hong}}, \bibinfo
  {author} {\bibfnamefont {Y.}~\bibnamefont {Chen}}, \bibinfo {author}
  {\bibfnamefont {Z.}~\bibnamefont {Sheng}}, \bibinfo {author} {\bibfnamefont
  {S.}~\bibnamefont {Wu}}, \bibinfo {author} {\bibfnamefont {Z.}~\bibnamefont
  {Wang}}, \bibinfo {author} {\bibfnamefont {B.}~\bibnamefont {Liang}},
  \bibinfo {author} {\bibfnamefont {W.-T.}\ \bibnamefont {Liu}}, \bibinfo
  {author} {\bibfnamefont {Z.}~\bibnamefont {Yuan}}, \bibinfo {author}
  {\bibfnamefont {Y.}~\bibnamefont {Wu}}, \bibinfo {author} {\bibfnamefont
  {Q.}~\bibnamefont {Mi}}, \bibinfo {author} {\bibfnamefont {Z.}~\bibnamefont
  {Liu}}, \bibinfo {author} {\bibfnamefont {J.}~\bibnamefont {Shen}},\ and\
  \bibinfo {author} {\bibfnamefont {S.}~\bibnamefont {Wu}},\ }\bibfield
  {title} {\bibinfo {title} {{Resolving and routing magnetic polymorphs in a 2D
  layered antiferromagnet}},\ }\href
  {https://doi.org/10.1038/s41563-024-02074-w} {\bibfield  {journal} {\bibinfo
  {journal} {Nature Materials}\ }\textbf {\bibinfo {volume} {24}},\ \bibinfo
  {pages} {226} (\bibinfo {year} {2025})}\BibitemShut {NoStop}%
\bibitem [{\citenamefont {Graham}\ \emph {et~al.}(2026)\citenamefont {Graham},
  \citenamefont {Wang}, \citenamefont {Nair}, \citenamefont {Mosina},
  \citenamefont {Watanabe}, \citenamefont {Taniguchi}, \citenamefont {Sofer},\
  and\ \citenamefont {Zhou}}]{Graham_2026_dataset}%
  \BibitemOpen
  \bibfield  {author} {\bibinfo {author} {\bibfnamefont {T.~K.~M.}\
  \bibnamefont {Graham}}, \bibinfo {author} {\bibfnamefont {Y.-X.}\
  \bibnamefont {Wang}}, \bibinfo {author} {\bibfnamefont {N.~R.}\ \bibnamefont
  {Nair}}, \bibinfo {author} {\bibfnamefont {K.}~\bibnamefont {Mosina}},
  \bibinfo {author} {\bibfnamefont {K.}~\bibnamefont {Watanabe}}, \bibinfo
  {author} {\bibfnamefont {T.}~\bibnamefont {Taniguchi}}, \bibinfo {author}
  {\bibfnamefont {Z.}~\bibnamefont {Sofer}},\ and\ \bibinfo {author}
  {\bibfnamefont {B.~B.}\ \bibnamefont {Zhou}},\ }\bibfield  {title} {\bibinfo
  {title} {Data pertaining to ``{S}pace-{C}harge-{L}imited van der {W}aals
  {S}pin {T}ransistor"},\ }\href {https://doi.org/10.5281/zenodo.20331365}
  {10.5281/zenodo.20331365} (\bibinfo {year} {2026})\BibitemShut {NoStop}%
\bibitem [{\citenamefont {Gross}\ \emph {et~al.}(2017)\citenamefont {Gross},
  \citenamefont {Akhtar}, \citenamefont {Garcia}, \citenamefont
  {Mart{\'{i}}nez}, \citenamefont {Chouaieb}, \citenamefont {Garcia},
  \citenamefont {Carr{\'{e}}t{\'{e}}ro}, \citenamefont
  {Barth{\'{e}}l{\'{e}}my}, \citenamefont {Appel}, \citenamefont {Maletinsky},
  \citenamefont {Kim}, \citenamefont {Chauleau}, \citenamefont {Jaouen},
  \citenamefont {Viret}, \citenamefont {Bibes}, \citenamefont {Fusil},\ and\
  \citenamefont {Jacques}}]{Gross2017}%
  \BibitemOpen
  \bibfield  {author} {\bibinfo {author} {\bibfnamefont {I.}~\bibnamefont
  {Gross}}, \bibinfo {author} {\bibfnamefont {W.}~\bibnamefont {Akhtar}},
  \bibinfo {author} {\bibfnamefont {V.}~\bibnamefont {Garcia}}, \bibinfo
  {author} {\bibfnamefont {L.~J.}\ \bibnamefont {Mart{\'{i}}nez}}, \bibinfo
  {author} {\bibfnamefont {S.}~\bibnamefont {Chouaieb}}, \bibinfo {author}
  {\bibfnamefont {K.}~\bibnamefont {Garcia}}, \bibinfo {author} {\bibfnamefont
  {C.}~\bibnamefont {Carr{\'{e}}t{\'{e}}ro}}, \bibinfo {author} {\bibfnamefont
  {A.}~\bibnamefont {Barth{\'{e}}l{\'{e}}my}}, \bibinfo {author} {\bibfnamefont
  {P.}~\bibnamefont {Appel}}, \bibinfo {author} {\bibfnamefont
  {P.}~\bibnamefont {Maletinsky}}, \bibinfo {author} {\bibfnamefont {J.-V.}\
  \bibnamefont {Kim}}, \bibinfo {author} {\bibfnamefont {J.~Y.}\ \bibnamefont
  {Chauleau}}, \bibinfo {author} {\bibfnamefont {N.}~\bibnamefont {Jaouen}},
  \bibinfo {author} {\bibfnamefont {M.}~\bibnamefont {Viret}}, \bibinfo
  {author} {\bibfnamefont {M.}~\bibnamefont {Bibes}}, \bibinfo {author}
  {\bibfnamefont {S.}~\bibnamefont {Fusil}},\ and\ \bibinfo {author}
  {\bibfnamefont {V.}~\bibnamefont {Jacques}},\ }\bibfield  {title} {\bibinfo
  {title} {{Real-space imaging of non-collinear antiferromagnetic order with a
  single-spin magnetometer}},\ }\href {https://doi.org/10.1038/nature23656}
  {\bibfield  {journal} {\bibinfo  {journal} {Nature}\ }\textbf {\bibinfo
  {volume} {549}},\ \bibinfo {pages} {252} (\bibinfo {year}
  {2017})}\BibitemShut {NoStop}%
\bibitem [{\citenamefont {Dirnberger}\ \emph {et~al.}(2023)\citenamefont
  {Dirnberger}, \citenamefont {Quan}, \citenamefont {Bushati}, \citenamefont
  {Diederich}, \citenamefont {Florian}, \citenamefont {Klein}, \citenamefont
  {Mosina}, \citenamefont {Sofer}, \citenamefont {Xu}, \citenamefont {Kamra},
  \citenamefont {Garc{\'{i}}a-Vidal}, \citenamefont {Al{\`{u}}},\ and\
  \citenamefont {Menon}}]{Dirnberger2023}%
  \BibitemOpen
  \bibfield  {author} {\bibinfo {author} {\bibfnamefont {F.}~\bibnamefont
  {Dirnberger}}, \bibinfo {author} {\bibfnamefont {J.}~\bibnamefont {Quan}},
  \bibinfo {author} {\bibfnamefont {R.}~\bibnamefont {Bushati}}, \bibinfo
  {author} {\bibfnamefont {G.~M.}\ \bibnamefont {Diederich}}, \bibinfo {author}
  {\bibfnamefont {M.}~\bibnamefont {Florian}}, \bibinfo {author} {\bibfnamefont
  {J.}~\bibnamefont {Klein}}, \bibinfo {author} {\bibfnamefont
  {K.}~\bibnamefont {Mosina}}, \bibinfo {author} {\bibfnamefont
  {Z.}~\bibnamefont {Sofer}}, \bibinfo {author} {\bibfnamefont
  {X.}~\bibnamefont {Xu}}, \bibinfo {author} {\bibfnamefont {A.}~\bibnamefont
  {Kamra}}, \bibinfo {author} {\bibfnamefont {F.~J.}\ \bibnamefont
  {Garc{\'{i}}a-Vidal}}, \bibinfo {author} {\bibfnamefont {A.}~\bibnamefont
  {Al{\`{u}}}},\ and\ \bibinfo {author} {\bibfnamefont {V.~M.}\ \bibnamefont
  {Menon}},\ }\bibfield  {title} {\bibinfo {title} {{Magneto-optics in a van
  der Waals magnet tuned by self-hybridized polaritons}},\ }\href
  {https://doi.org/10.1038/s41586-023-06275-2} {\bibfield  {journal} {\bibinfo
  {journal} {Nature}\ }\textbf {\bibinfo {volume} {620}},\ \bibinfo {pages}
  {533} (\bibinfo {year} {2023})}\BibitemShut {NoStop}%
\bibitem [{\citenamefont {Wang}\ \emph {et~al.}(2017)\citenamefont {Wang},
  \citenamefont {Kim}, \citenamefont {Kim}, \citenamefont {Kim}, \citenamefont
  {Hwang}, \citenamefont {Cho}, \citenamefont {Shin}, \citenamefont {Das},
  \citenamefont {Kim}, \citenamefont {Kalinin}, \citenamefont {Kim},
  \citenamefont {Yang},\ and\ \citenamefont {Noh}}]{Wang2017d}%
  \BibitemOpen
  \bibfield  {author} {\bibinfo {author} {\bibfnamefont {L.}~\bibnamefont
  {Wang}}, \bibinfo {author} {\bibfnamefont {R.}~\bibnamefont {Kim}}, \bibinfo
  {author} {\bibfnamefont {Y.}~\bibnamefont {Kim}}, \bibinfo {author}
  {\bibfnamefont {C.~H.}\ \bibnamefont {Kim}}, \bibinfo {author} {\bibfnamefont
  {S.}~\bibnamefont {Hwang}}, \bibinfo {author} {\bibfnamefont {M.~R.}\
  \bibnamefont {Cho}}, \bibinfo {author} {\bibfnamefont {Y.~J.}\ \bibnamefont
  {Shin}}, \bibinfo {author} {\bibfnamefont {S.}~\bibnamefont {Das}}, \bibinfo
  {author} {\bibfnamefont {J.~R.}\ \bibnamefont {Kim}}, \bibinfo {author}
  {\bibfnamefont {S.~V.}\ \bibnamefont {Kalinin}}, \bibinfo {author}
  {\bibfnamefont {M.}~\bibnamefont {Kim}}, \bibinfo {author} {\bibfnamefont
  {S.~M.}\ \bibnamefont {Yang}},\ and\ \bibinfo {author} {\bibfnamefont
  {T.~W.}\ \bibnamefont {Noh}},\ }\bibfield  {title} {\bibinfo {title}
  {{Electronic‐Reconstruction‐Enhanced Tunneling Conductance at Terrace
  Edges of Ultrathin Oxide Films}},\ }\href
  {https://doi.org/10.1002/adma.201702001} {\bibfield  {journal} {\bibinfo
  {journal} {Advanced Materials}\ }\textbf {\bibinfo {volume} {29}},\ \bibinfo
  {pages} {1702001} (\bibinfo {year} {2017})}\BibitemShut {NoStop}%
\end{thebibliography}

\setcounter{section}{0}

\section{Acknowledgments}
The authors thank Hua Chen, Vsevolod Belosevich and Zicheng Ying for valuable discussions. B.B.Z. and T.K.M.G. acknowledge support from the Department of Energy Early Career Program under award number DE-SC0024177 for device fabrication, electrical transport and magnetic imaging measurements. B.B.Z., T.K.M.G. and Y.-X.W. acknowledge support from the National Science Foundation award DMR-2047214 for development of the scanning NV microscope. Z.S. was supported by project LUAUS25268 from Ministry of Education Youth and Sports (MEYS) and by the project Advanced Functional Nanorobots (reg. No. CZ.02.1.01/0.0/0.0/15\_003/0000444 financed by the EFRR). K.W. and T.T. acknowledge support from the JSPS KAKENHI (Grant Numbers 21H05233 and 23H02052), the CREST (JPMJCR24A5), JST and World Premier International Research Center Initiative (WPI), MEXT, Japan.

\section{Data availability} The data that support the findings of this article are openly available \cite{Graham_2026_dataset}.


%



\section*{End Matter}

\begin{figure*}[t]
\includegraphics[scale=1]{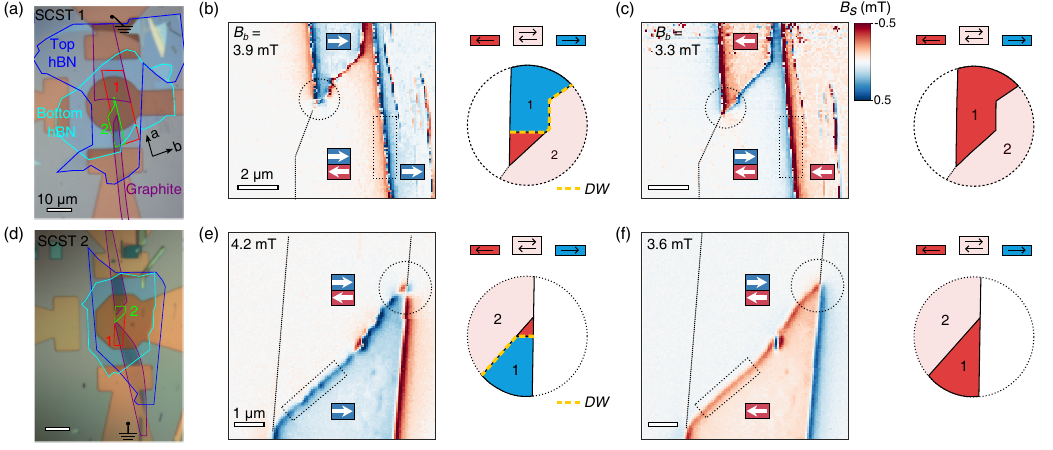}
\caption{\label{fig:6} Resolving the antiphase states of bilayer CrSBr. (a) Optical image of SCST 1, corresponding to the device presented in the main text. (b) Stray field $B_S$ image of the monolayer-bilayer interface when a domain wall exists between the monolayer and bilayer. The monolayer is identified as $\rightarrow$ by its negative $B_S$, from which we deduce that the bilayer is \RL, denoting the bilayer's top and bottom layer magnetization, respectively. The domain wall (DW) is revealed by the bumpy monolayer-bilayer edge (box) and by a perturbation at the monolayer corner (circle) where the domain wall cuts the corner (see cartoon). (c) Stray field image of SCST 1 when no domain wall is present after flipping the monolayer to $\leftarrow$. The same edge is now smooth, and the corner is unblemished. (d) Optical image of a second device, SCST 2. (e,f) Images when a domain wall is present (e) and absent (f) in SCST 2. Similar telltale features as SCST 1 are observed, enabling determination of the bilayer state as \RL.}
\end{figure*}

\paragraph*{Appendix A: Experimental methods.---} The gated SCST and sf-MTJ devices were fabricated by a multistep procedure. First, Ti/Au/Pt electrical contacts (see Fig. \ref{fig:6}a and \ref{fig:6}d) were fabricated onto \Si wafers using photolithography and electron beam evaporation, and then annealed for 3 hours at $350^\circ$C in forming gas (Ar/H$_2$, 95:5). Next, an hBN flake (bottom hBN), serving as the gate dielectric, was transferred onto the backgate electrode using polycarbonate (PC)-assisted dry transfer. A second forming gas anneal was performed to remove transfer residues. The thickness of the hBN gate dielectric for SCST 1 (main text) was measured by atomic force microscopy to be 12 nm. Assuming parallel plate capacitance, we estimate an ideal doping efficiency of $1.7\times10^{12}$ $e$ cm$^{-2}$ V$^{-1}$, which is distributed between trapped and free carriers. Moreover, the actual gate-induced doping distribution is likely to be nonuniform in the low free carrier regime owing to lateral potential gradients due to charge trapping.

CrSBr flakes were exfoliated onto \Si (285 nm oxide) at 60$^\circ$C in an argon glovebox using Scotch Magic Green tape. For the SCST, flakes containing a monolayer–bilayer step were identified by optical contrast, with each additional CrSBr layer reducing the reflected red intensity by $\approx$10\% relative to the background \cite{Note1}. Electrical contacts bridging the CrSBr and prepatterned electrodes were prepared from few-layer graphene (graphite) exfoliated at 70$^\circ$C. Two graphite strips were sequentially picked up by a $\sim$10 nm-thick hBN flake (top hBN), ensuring a separation of a few microns between the graphite strips. The top hBN with graphite electrodes was then carefully aligned onto CrSBr such that one electrode touched the bilayer and the other the monolayer and brought into contact at 90$^\circ$C to pick up the CrSBr flake. The entire stack was finally placed onto the preprepared metal contacts, with the CrSBr centered on the backgate and the graphite electrodes touching the source/drain contacts.

\begin{figure}[h]
\includegraphics[scale=1]{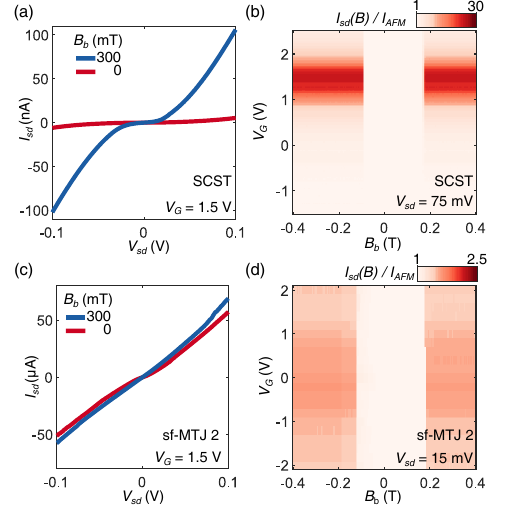}
\caption{\label{fig:5} Comparison of the SCST and sf-MTJ transport properties. (a) Current $I_{sd}$ versus bias $V_{sd}$ for the SCST at $B_b = 300$~mT (FM) and 0~mT (AFM) on a linear scale. (b) Normalized conductance $I_{sd}(B) / I_{AFM}$, where $I_{AFM}$ is the current in the AFM state, as a function of gate $V_G$ and field $B_b$ along the $b$-axis. (c) $I_{sd}$ versus $V_{sd}$ curve for the FM and AFM states of sf-MTJ 2. (d) Normalized conductance $I_{sd}(B) / I_{AFM}$ versus $V_G$ and $B_b$ for sf-MTJ 2. The MR of the SCST is tunable with $V_G$ and can achieve significantly higher values than the sf-MTJ.}
\end{figure}

The sf-MTJs were similarly assembled, placing the top and bottom graphite electrodes on opposite sides of a bilayer region while minimizing their overlap area. The bottom electrode is chosen to be monolayer graphene to minimize screening of the metallic backgate. In Fig.~ \ref{fig:1}e, the dip in current for the sf-MTJ near $V_G = 0$ is due to the Dirac point of the bottom electrode. Final devices were cleaned in sequential baths of chloroform, acetone and isopropanol to remove PC residues.

Electrical transport and scanning NV magnetic imaging were performed inside a customized closed cycle cryostat (attoDRY 2200) with natural isotope abundance, $\langle100\rangle$ diamond probe tips from QZabre with a nominal sensitivity of $1.9\;\mu \mathrm{T}/\sqrt{\mathrm{Hz}}$. The NV center implantation depth, according to the manufacturer, is $\approx$10 nm, while the standoff distance, setting the spatial resolution, was determined from stray field linecuts of the CrSBr edge to be $\approx$50 nm \cite{Note1}. The CrSBr devices were oriented to align their $b$-axes with the in-plane direction of the NV center tip. For electrical measurements presented in the main text, the external field was applied solely along the $b$-axis ($B_b$), while for NV imaging, an out-of-plane component ($B_c$ along the $c$-axis, hard axis) was added to orient the total field parallel to the NV center axis, with $\arctan(B_c/B_b) = 35^\circ$. Due to the strong magnetic anisotropy of CrSBr, its in-plane magnetization and spin-flip field are not significantly changed by this additional out-of-plane field (see SM for field-angle-dependent measurements \cite{Note1}). Hence, for the NV center images, we quote only the $B_b$ component of the total applied field. All data presented in the main text were taken at $T = 2$~K. 


\paragraph*{Appendix B: Additional transport data---}
Figure \ref{fig:5}a compares the device current $I_{sd}$ obtained in the AFM and FM states of the SCST on a linear scale, highlighting its drastic MR. The evolution of the SCST's conductance versus field $B_b$ along the $b$-axis and gate voltage $V_G$ is shown in Fig. \ref{fig:5}b. Analogous data for sf-MTJ 2 (a different device from the main text) is presented in Fig. \ref{fig:5}c,d, which corroborates the relatively low, gate-independent MR of the sf-MTJ. The spin-filter efficiency of bilayer CrSBr \cite{Boix-Constant2022} is lower than CrI$_3$ \cite{Song2018a} due to the former's weaker exchange splitting and higher vertical conductivity.

\paragraph*{Appendix C: Determination of the bilayer antiphase state.---} Due to the finite sample-to-NV-center separation ($\sim$50 nm) \cite{Note1}, the tiny difference in stray field between the antiphase states of the bilayer (\RL{} versus \LR{}) cannot be directly resolved \cite{Wang2025a}. However, for the SCST, the bilayer's lateral connection to a monolayer region, whose uncompensated magnetization is easily measured, allows the bilayer's state to be deduced from whether a domain wall is present against the monolayer. As shown in Fig.~\ref{fig:6} for two different SCST devices, the edge between bilayer and monolayer appears ragged in magnetic imaging when a domain wall is present in their shared layer (e.g., identifying the bilayer as \RL{} when the monolayer is $\rightarrow$), versus smooth when a domain wall is absent. Moreover, as highlighted in the circled regions, the domain wall, if present, lowers its energy by cutting the corner where the monolayer meets the bilayer at an acute angle, leading to a sign switch of the stray magnetic field at the monolayer corner (see SM for zoomed-in data, magnetization reconstruction and comparisons to stray field simulations \cite{Note1}). This direct visualization of the bilayer antiphase state allows us to identify which layer of the bilayer flips at each AFM-to-FM or FM-to-AFM transition on the forward and backward field sweeps, elucidating the magnetic training effect and its manifestation in the device conductance (Fig. \ref{fig:3}).


\clearpage


\end{document}